\documentclass[journal]{IEEEtran}

\usepackage{amsmath,amsfonts,amsthm} 
\usepackage{graphicx}
\usepackage{booktabs,makecell}
\usepackage[table]{xcolor}
\usepackage{flushend}
\usepackage{amsmath,graphicx,amsfonts,amssymb,fixmath,psfrag,url,booktabs,multirow,makecell}
\usepackage{comment}
\usepackage{cite}
\usepackage{xcolor}
\usepackage{color}
\usepackage{graphicx}
\usepackage{changes}
\usepackage{color}
\usepackage{graphicx}
\usepackage{paralist}
\usepackage{flexisym}
\usepackage{subcaption}
\usepackage{amsmath,graphicx,amsfonts,amssymb,fixmath,psfrag,url,booktabs,multirow}
\usepackage{comment}
\usepackage{cite}
\usepackage{xcolor}
\usepackage{mwe}
\usepackage{color}
\usepackage{graphicx}
\usepackage{mathtools}
\usepackage[utf8]{inputenc}
\usepackage{fancyhdr} 
\usepackage{enumerate}
\usepackage{mathtools, cuted}
\usepackage{dblfloatfix}
\usepackage{diagbox}
\usepackage{tabularx}
\usepackage[many]{tcolorbox}
\usepackage{adjustbox}
\usepackage{framed}

\newtheorem{proposition}{Proposition}
\newtheorem{corollary}{Corollary}
\newtheorem{theorem}{Theorem}
\newtheorem{definition}{Definition}
\newtheorem{remark}{Remark}

\renewcommand{\a}{\mathbold{a}}
\renewcommand{\b}{\mathbold{b}}

\newcommand{\p}{\mathbold{p}}
\newcommand{\x}{\mathbold{x}}
\newcommand{\y}{\mathbold{y}}
\newcommand{\w}{\mathbold{w}}
\newcommand{\z}{\mathbold{z}}

\newcommand{\A}{\mathbold{A}}
\newcommand{\B}{\mathbold{B}}

\newcommand{\D}{\mathbold{D}}
\newcommand{\E}{\mathbold{E}}

\renewcommand{\H}{\mathbold{H}}
\newcommand{\I}{\mathbold{I}}
\newcommand{\K}{\mathbold{K}}
\renewcommand{\L}{\mathbold{L}}

\renewcommand{\P}{\mathbold{P}}
\newcommand{\R}{\mathbold{R}}

\newcommand{\U}{\mathbold{U}}
\newcommand{\W}{\mathbold{W}}
\newcommand{\X}{\mathbold{X}}
\newcommand{\Y}{\mathbold{Y}}

\newcommand{\ie}{i.e.,~}
\newcommand{\eg}{e.g.,~}

\newcommand{\LAM}{\mathbf{\Lambda}}

\newcommand{\transp}{\top}
\newcommand{\herm}{{\mathsf{H}}}

\newcommand{\Gr}{\mathcal{G}}
\newcommand{\Ed}{\mathcal{E}}
\newcommand{\Exp}{\mathbb{E}}
\newcommand{\Vx}{\mathcal{V}}

\newcommand{\bSigma}{\mathbf{\Sigma}}    
\newcommand{\trace}{\text{tr}}           
\newcommand{\Scal}{\mathcal{S}}
\newcommand{\bPsi}{\mathbf{\Psi}}

\newcommand{\GFT}[1]{\textrm{GFT}\hspace{-.0mm}\{#1\}}

\newcommand{\DFT}[1]{\textrm{DFT}\hspace{-.0mm}\{#1\}}
\newcommand{\JFT}[1]{\textrm{JFT}\hspace{-.0mm}\{#1\}}

\newcommand{\Rbb}{\mathbb{R}} 

\newcommand{\eps}{\boldsymbol{\varepsilon}} 
\renewcommand{\qed}{\hfill\blacksquare}
\newcommand{\rev}{\textcolor{black}}

\newtcolorbox{annotatedbox}[1][]{%
  enhanced jigsaw,
  sharp corners,
  colback=white,
  float=htb,
  boxsep=0pt,
  #1%
}

\DeclareMathOperator*{\minimize}{minimize}
\DeclareMathOperator*{\argmin}{arg\,min}

\begin{document}
\title{Forecasting Network Processes with VARMA Recursions on Graphs}
\title{Forecasting Time Series with VARMA\\ Recursions on Graphs}
%
%
%


\newcommand{\andreas}[1]{{\color[rgb]{.0,.1,.7}{#1}}}
\newcommand{\elvin}[1]{{\color[rgb]{.0,.5,.0}{#1}}}

\author{Elvin Isufi~\IEEEmembership{Student Member,~IEEE,}, Andreas Loukas, Nathanael Perraudin and Geert Leus~\IEEEmembership{Fellow,~IEEE,}
\thanks{Parts of this paper were presented at the \emph{2017 IEEE Asilomar conference} \cite{LoukasPred}.
E. Isufi and G. Leus are with the EEMCS Faculty, Delft University of Technology, 2628 CD, Delft, The Netherlands.
A. Loukas is with the LTS2 Laboratory, Swiss Federal Institute of Technology in Lausanne (EPFL), Switzerland.
N. Perraudin is with the Swiss Data Science Center, ETH Zurich and EPFL, Switzerland.
E-mails: \{e.isufi-1; g.j.t.leus\}@tudelft.nl; andreas.loukas@epfl.ch; n.perraudin@sdsc.ethz.ch}} 

\maketitle

\begin{abstract}

Graph-based techniques emerged as a choice to deal with the dimensionality issues in modeling multivariate time series. However, there is yet no complete understanding of how the underlying structure could be exploited to ease this task. 
This work provides contributions in this direction by considering the forecasting of a process evolving over a graph. We make use of the (approximate) time-vertex stationarity assumption, i.e., time-varying graph signals whose first and second order statistical moments are invariant over time and correlated to a known graph topology. The latter is combined with VAR and VARMA models to tackle the dimensionality issues present in predicting the temporal evolution of multivariate time series.

We find out that by projecting the data to the graph spectral domain: (\emph{i}) the multivariate model estimation reduces to that of fitting a number of uncorrelated univariate ARMA models and (\emph{ii}) an optimal low-rank data representation can be exploited so as to further reduce the estimation costs.
In the case that the multivariate process can be observed at a subset of nodes, the proposed models extend naturally to Kalman filtering on graphs allowing for optimal tracking. Numerical experiments with both synthetic and real data validate the proposed approach and highlight its benefits over state-of-the-art alternatives. 
\end{abstract}


\section{Introduction}
\label{sec.intro}

Forecasting multivariate processes is a central topic in signal processing \cite{box2015time,lutkepohl2005new}. Especially in the high-dimensional setting, the forecasting of \emph{unstructured data} is considered a complex task, both from an estimation and computational perspective --indeed, most works in this direction stay within the limit of a few dozen time series. To cope with the dimensionality challenge, particularly when the number of observations is smaller than the number of time series, typical solutions involve factor models \cite{connor1995three,lam2011factor}, shrinkage estimators \cite{meucci2009risk,ledoit2004well}, and low-rank data representations \cite{lutkepohl2005new}.

Recently, forecasting has been considered for \emph{graph-structured} processes, i.e., time-varying signals living on the nodes of graphs~\cite{mei2017signal,di2016adaptive,di2017adaptive,Qiu2017,romero2016kernel}. Here, given a graph and a set of time series, the aim is to exploit the underlying structure for improving the prediction accuracy of the considered model. The core idea is that the additional information, carried by the graph, can enhance our ability to make relevant predictions. The graph basically aids to restrict the degrees of freedom (DoFs) of the model in a manner faithful to the problem at hand. Depending on the scenario, these works exploit different signal priors w.r.t. the graph to formulate the prediction task. 

The first to consider prediction in the graph context is the work of~\cite{mei2017signal}, which used the vector autoregressive (VAR) recursions to learn the underlying topology and then used this structure to forecast future values. Their findings rely on a temporal interpretation of edges. More specifically, it is assumed that the information requires a unit time step to reach adjacent nodes. 
Coupling the duration of graph interactions and the fixed temporal resolution in such a manner may be beneficial for graph learning purposes, but negatively affects forecasting since it ignores hidden correlations between nodes.

Relevant work has also been done in \emph{tracking} time-varying signals on graphs\footnote{For the sake of clarity, the reader should distinguish between forecasting and tracking a time-varying signal. In forecasting, $x_t$ is predicted based on historical data $\{x_0, \ldots, x_{t-1}\}$, while in tracking noisy measurements at time $t$ are also used.}. In the adaptive approaches~\cite{di2016adaptive,di2017adaptive}, the prediction/tracking of time-varying graph signals comes as a byproduct of the presented adaptive estimation strategies. Here, the authors exploit a low-rank representation of the signal, due to the bandlimitedness in the graph spectral domain, to track slow signal variations over time. 

A different approach for network process tracking is followed in \cite{Qiu2017,romero2016kernel}. Here, a virtual graph extrapolated to the temporal dimension is considered and the multivariate time series is treated as a single time-invariant (extended) graph signal on this larger graph. Then, under priors such as diffusion or smoothness, the tracking task is rephrased as an interpolation problem. To ease the excessive computational burden, gradient descent algorithms \cite{Qiu2017} or Kalman-based recursions \cite{romero2016kernel} are invoked.

Almost all the above works treat the time series as the evolution of a noise-corrupted deterministic process. However, in many practical situations including opinion blogs, temperature sensor networks, financial markets, or brain networks, the time series has a strong stochastic component. The latter requires models able to capture the stochasticity along both the graph and time dimension and allow the short time prediction from few training samples. Some earlier contributions have considered stochasticity in time, see e.g., \cite{mei2017signal,di2017adaptive}, yet leaving unexplored the stochasticity along the graph dimension.

This work takes one step further and treats time series as wide-sense stationary (for short stationary) statistical processes w.r.t. both the graph and temporal domains. While temporal stationarity preserves the process statistics w.r.t. translations in time, the considered joint stationarity hypothesis also preserves these statistics w.r.t. graph localization. \rev{A typical example comprises temperature measurements in sensor networks. The graph stationarity implies that neighboring sensors yield in expectation a constant value. The joint stationary, on the other hand, considers the expected constant value to be also preserved in consecutive time instances}\footnote{We remark that although perfect stationarity either in the time or graph domain is difficult to be observed, several real data sets \cite{Girault2015,Perraudin2016,Marques2016,loukas2016stationary} have shown some degree of graph/time stationarity that can be exploited for prediction.}.

Continuing our prior works \cite{loukas2016stationary,Isufi2016separable}, we exploit the (approximate) time-vertex stationarity of graph time series and extend classical VAR and vector autoregressive moving average (VARMA) recursions for modeling and predicting time-varying processes on graphs. Specifically, the contributions of this work are:
\begin{enumerate}
\vskip1mm
\item We propose VAR and VARMA models for forecasting time series on graphs. These recursions extend the non-causal models of \cite{loukas2016stationary} to causal ones and generalize the approach of \cite{mei2017signal} to $(i)$ VAR recursions that do not impose a temporal interpretation of the edges, and $(ii)$ VARMA recursions where the MA part allows to model time series with a more complete structure. 
\vskip1mm
\item We propose a model fitting method that makes use of the data structure and stationarity. 
Specifically, we separate the multivariate process into uncorrelated univariate time series (one per graph frequency), hence tackling the dimensionality issue present in VARMA modeling. The latter enables us to use well-established univariate techniques for estimating the model coefficients from limited training data. We further introduce an optimal low-rank data representation, which intrinsically makes use of the graph to further reduce the computational overhead of the model estimation.
\vskip1mm
\item We introduce a sub-graph tracker to estimate the time series on a subset of nodes. The latter makes use the prediction at time $t$ along with graph sub-sampled measurements to estimate the attributes on the nodes of interest. We show a direct connection between the proposed approach and Kalman filtering (KF) on graphs \cite{isufi2017observing} that allows a direct characterization of the sub-graph tracker performance. 
%
\end{enumerate}
The paper makes an additional minor contribution. It generalizes the separable two-dimensional time-vertex ARMA filter of \cite{Isufi2016separable} to a non-separable version and shows that the latter encompasses other state-of-the-art time-vertex recursions on graphs as the ones presented in \cite{mei2017signal,Isufi2016Dim2} and can be used for distributed predictors.

To validate the joint stationarity hypothesis and the proposed models, we consider three scenarios: $(i)$ {the Molene weather} data set\footnote{\rev{Data publicly available at \texttt{https://donneespubliques.\\meteofrance.fr/donnees\_libres/Hackathon/RADOMEH.tar.gz}.}}; the NOAA U.S. temperature data set \cite{nooaDataset} $(ii)$ a dynamic mesh data set containing a walking dog; and $(iii)$ an epidemic spreading scenario over a flight transportation network. The obtained results corroborate the relevance of graph VARMA models for forecasting stochastic time series supported on graphs and show that the proposed approaches outperform both classical univariate and multivariate ARMA techniques and state-of-the-art graph-based techniques. 


\vskip1mm
\noindent\textbf{Notation.} The scalar $a_i$ indicates the $i$th entry of vector $\a$, and similarly the scalar $A_{i,j}$ indicates the $(i,j)$th entry of matrix $\A$. When needed these entries will also be denoted as $[\a]_i$, or $[\A]_{ij}$. $\A^\transp$, $\A^*$ and $\A^{\herm}$ are respectively the transpose, the element-wise conjugate, and the Hermitian of $\A$. $\|\a\|_p$ and $\|\A\|_p$ indicate respectively the $p$-th norm of $\a$ and $\A$, whereas $\|\A\|_F$ is the Frobenius norm of matrix $\A$. The trace of $\A$ is denoted as $\trace(\A)$ and $\A \otimes \B$ is the Kronecker product of the matrices $\A$ and $\B$. $\a = \text{diag}(\A)$ is the operation that stores the diagonal elements of $\A$ into the vector $\a$, and $\A = \text{diag}(\a)$ is a diagonal matrix containing the vector $\a$ in the main diagonal. \rev{The operation $\a = \text{vec}(\A)$ concatenates the columns of $\A$ in $\a$} and $\text{vec}^{-1}(\cdot)$ is the inverse operation. The $N \times N$ identity matrix is denoted as $\I_N$, the $N \times 1$ all ones vector as $\mathbf{1}_N$, and the $N \times 1$ all zeros vector as $\mathbf{0}_N$. \rev{For a set $\Scal \subseteq \{1, \ldots, N\}$, $|\Scal|$ is the set cardinality, and $\D_\Scal$ is an $N \times N$ diagonal matrix with $[\D_\Scal]_{ii} = 1$ if $i \in \Scal$ and zero otherwise.} To avoid confusion, we refer to \emph{causality} in the classical sense \cite{lutkepohl2005new}, i.e., an event happening at time $t$ is influenced only by events prior to $t$ independently of the graph dimension. We will refer to the notion of causality used by \cite{mei2017signal} as \emph{restricted causality}, being a subclass of the former.

\vskip1mm
\noindent\textbf{Paper organization.} Section~\ref{sec.back} contains the background information. Section~\ref{sec.mod} formulates the prediction problem in the graph signal processing (GSP) perspective and develops the optimal predictor. Section~\ref{sec.fit} focuses on the model parameter design.
The subgraph tracker is introduced and analyzed in Section~\ref{sec.subset}. Section~\ref{sec.numRes} contains the numerical results and the paper conclusions are drawn in Section~\ref{sec.concl}. 

\section{Preliminaries}
\label{sec.back}

This section starts with the fundamentals of graph and time-vertex signal processing. Then, it proceeds with the two-dimensional time-vertex ARMA models and the definition of joint stationarity that will be used in the rest of the paper.

\subsection{Graph Signal Processing}
\label{subsec_gsp}

Consider a weighted undirected graph $\Gr = (\Vx,\Ed,\W_\Gr)$, where $\Vx$ is the set of $N$ nodes (vertices), $\Ed$ is the edge set, and $\W_\Gr$ is the weighted adjacency matrix with $[\W_\Gr]_{ij} = [\W_\Gr]_{ji} > 0$ if $(i,j) \in \Ed$ and zero otherwise. A graph signal is a mapping from the vertex set to the set of complex numbers, \ie x $: \Vx \to \mathbb{C}$. For convenience we represent a graph signal by the vector $\x = [x_1,\ldots , x_N]^{\transp}$, where $x_i$ is the value related to node $i$. The graph Fourier transform (GFT) of $\x$ is
\begin{equation}
\label{eq.gft}
\GFT{\x} = \U_\Gr^\herm\x,
\end{equation}
where $\U_\Gr$ is the unitary eigenvector matrix of the discrete Laplacian matrix $\L_\Gr = \text{diag}(\W_\Gr\mathbf{1}_N) - \W_\Gr = \U_\Gr\LAM_\Gr\U_\Gr^\herm$. The diagonal matrix $\LAM_\Gr$ contains the eigenvalues on its main diagonal, often referred to as the graph frequencies \cite{Shuman2013}. The GFT allows filtering on graphs in the graph frequency domain. The filtered version $\y$ of $\x$ with the graph filter $h(\L_\Gr) \in \mathbb{R}^{N\times N}$ corresponds to an element-wise multiplication in the spectral domain, \ie
\begin{equation}
    \y = h(\L_\Gr)  \x \triangleq \U_\Gr h(\LAM_\Gr) \U_\Gr^\herm \, \x, 
\end{equation}
with matrix $h(\L_\Gr)$ obtained by applying the scalar function $h: [0, \lambda_{\textit{max}}] \rightarrow \Rbb$ on the eigenvalues of $\L_\Gr$. Here, $h(\LAM_\Gr)$ is diagonal and is referred to as the graph filter frequency response. See \cite{taubin2000geometric,Shuman2013,sandryhaila2013discrete} for more details on GSP and \cite{shuman2011chebyshev,Shi2015, segarra2017optimal,isufi2017autoregressive} for graph filters.

\subsection{Time-Vertex Signal Processing}
\label{subsec_gtsp}

For a time-varying graph signal $\x_t$, let $\X_{1:T} = [\x_1, \x_2, \ldots, \x_T] \in \mathbb{R}^{N \times T}$ be the $N \times T$ matrix that collects $T$ successive realizations. Let also $\x_{1:T} = \text{vec}(\X_{1:T})$ denote the vectorized form of $\X_{1:T}$. From \cite{grassi2017time}, the joint (time-vertex) Fourier transform (JFT) of $\X_{1:T}$ is
\begin{align}
\label{eq.jft_def}
    \JFT{\X_{1:T}} = \U_\Gr^{\herm} \X_{1:T} {\U}_{\mathcal{T}}^*,
\end{align}
where, once again $\U_\Gr$ is the graph Laplacian eigenvector matrix, \rev{while ${\U}_{\mathcal{T}}^* \in \mathbb{C}^{T\times T}$ is the complex conjugate} of the discrete Fourier transform (DFT) matrix. $\U_{\mathcal{T}}$ can also be interpreted as the unitary eigenvector matrix of the circulant time-shift operator $\L_{\mathcal{T}} = \U_{\mathcal{T}} \, \LAM_{\mathcal{T}} \, \U_{\mathcal{T}}^\herm$\rev{
\begin{equation}
[\LAM_{\mathcal{T}}]_{tt} = e^{-\j \omega_t}, \quad \text{and}\quad \omega_t = \frac{2\pi (t-1)}{T}
\end{equation}
for $\j = \sqrt{-1}$ and $t = 1, \ldots, T$}. In a vectorized form, \eqref{eq.jft_def} becomes $\JFT{\x_{1:T}} = \U_{\mathcal{J}}^\herm\x$, with $\U_{\mathcal{J}} = \U_{\mathcal{T}}\otimes\U_\Gr$ being a unitary matrix.

Similar to the GFT, the JFT allows a \emph{joint} time-vertex filtering of $\x_{1:T}$ in the joint time-vertex frequency domain. The latter is given by the eigendecomposition of the extended Laplacian operator $\L_{\mathcal{J}} = \I_N \otimes \L_\Gr + \L_{\mathcal{T}}\otimes \I_N$ \cite{grassi2017time}. The joint filtered version $\y_{1:T}$ of $\x_{1:T}$ with a joint filter $h(\L_{\mathcal{J}})$ corresponds again to an element-wise multiplication, but now in the joint frequency representation, i.e.,
\begin{equation}  \label{eq:def_joint_filtering}
    \y_{1:T} = h(\L_{\mathcal{J}}) \x_{1:T} \triangleq \U_{\mathcal{J}} \, h(\LAM_\Gr, \LAM_{\mathcal{T}} ) \, \U_{\mathcal{J}}^\herm \x_{1:T}. 
\end{equation} 
{Here, $h(\LAM_\Gr,\LAM_{\mathcal{T}})$ is an $NT\times NT$ diagonal matrix called the joint time-vertex frequency response with $k$th diagonal element $[h(\LAM_\Gr,\LAM_{\mathcal{T}})]_{kk} = h(\lambda_n,e^{\j\omega_t})$ and $k = N(t-1)+n$ for $n = 1, \ldots, N$ and $t = 1, \ldots, T$.} By means of inverse vectorization, we write $\Y_{1:T} = \textnormal{vec}^{-1}(\y_{1:T})$ and the $t$th column of $\Y_{1:T}$, $\y_t$, is the filtered signal at time $t$. 

\vspace{2mm}\noindent\textbf{Time-vertex ARMA filters.} An efficient way to (distributively) implement \eqref{eq:def_joint_filtering} is by extending the two-dimensional ARMA filters \cite{Isufi2016separable} to the more general form
\begin{equation}
\label{eq.2dARMA}
\y_t + \sum_{p = 1}^P\sum_{l = 0}^{L_p}\psi_{l,p}\L_\Gr^l\y_{t-p}= \sum_{q = 0}^Q\sum_{k=0}^{K_q}\varphi_{k,q}\L_\Gr^k\x_{t-q},
\end{equation}
where $P, L_p, Q, K_q$ are positive scalars and $\psi_{l,p}, \varphi_{k,q}$ denote complex coefficients. In computing the filter output $\y_t$, recursion \eqref{eq.2dARMA} considers a linear combination of the current and past graph signal realizations $\{\x_{t}, \ldots, \x_{t-Q}\}$ and of the previous outputs $\{\y_{t-1}, \ldots, \y_{t-P}\}$. This temporal shifting of the filter input-output provides a filtering in the temporal domain. The graph filtering is provided by shifting each temporal realization $\x_{t-\tau}$ ($\y_{t-\tau}$) in the graph dimension using powers of $\L_\Gr$.

To see more formally the two-dimensional filtering effect of \eqref{eq.2dARMA}, we first apply the DFT and then the GFT on both sides of \eqref{eq.2dARMA} to obtain the per time-vertex frequency relation
\begin{align}
\label{eq.2dARMA_freq}
\left(\!\!1 \!+\!\! \sum_{p = 1}^P\sum_{l = 0}^{L_p}\psi_{l,p}\lambda_n^le^{-\j \omega_t p}\right)\hat{y}_{n}(e^{\j\omega_t}) &= \notag \\
&\hspace{-25mm}\sum_{q = 0}^Q\sum_{k=0}^{K_q}\varphi_{k,q}\lambda_n^le^{-\j \omega_t q}\hat{x}_{n}(e^{\j\omega_t}),
\end{align}
where $\hat{y}_{n}(e^{\j\omega_t})$ ($\hat{x}_{n}(e^{\j\omega_t})$) is the $n$th entry of $\GFT{\DFT{\y_t}}$ ($\GFT{\DFT{\x_t}}$). Then, for stable filter coefficients $\psi_{l,p}$, \ie coefficients that abide to 
%
%
\begin{equation}
\label{eq.2dARMA_stable}
1 + \sum_{p = 1}^P\sum_{l = 0}^{L_p}\psi_{l,p}\lambda_n^le^{-\jmath \omega_t p} \neq 0, \quad \forall~ n = 1, \ldots, N,
\end{equation}
\eqref{eq.2dARMA} implements the two-dimensional time-vertex frequency response
\begin{equation}
\label{eq.2dARMA_fresp}
h(\lambda_n,e^{\j \omega_t}) = \frac{\sum_{q = 0}^Q\sum_{k=0}^{K_q}\varphi_{k,q}\lambda_n^ke^{-\j \omega_t q}}{1 + \sum_{p = 1}^P\sum_{l = 0}^{L_p}\psi_{l,p}\lambda_n^le^{-\j \omega_t p}}.
\end{equation}
It is easy to see that the above frequency response is a joint time-vertex filter. 
Similarly to its separable counterpart \cite{Isufi2016separable}, recursion \eqref{eq.2dARMA} considers polynomial shifting in the graph dimension. Thus, from the locality of $\L_\Gr$ \cite{Shuman2013}, and since $\L_\Gr^k\x_{t-q} = \L_\Gr(\L_\Gr^{k-1}\x_{t-q})$ \cite{shuman2011chebyshev,segarra2017optimal}, recursion \eqref{eq.2dARMA} enjoys the same distributed implementation as \cite{Isufi2016separable}.

We conclude this section with the following observation.

\begin{remark}[Model generality] Recursion \eqref{eq.2dARMA} encompasses some of the state-of-the-art two-dimensional time-vertex models. Specifically:
\begin{itemize}
\item for $\psi_{l,p} = \psi_la_p$ and $\varphi_{p,q} = \varphi_pb_q$ it specializes to the separable case~\cite{Isufi2016separable};
\item for $\psi_{l,p} = 0$ $\forall l, p$, it boils down to the two-dimensional finite impulse response time-vertex filter~\cite{Isufi2016Dim2};
\item for $L_p = p$, $\sum_{k=0}^{K_0}\varphi_{k,0}\L_\Gr^k = \I_N$ and $\varphi_{k,q} = 0$ otherwise, it reduces to the model of \cite{mei2017signal}.
\end{itemize}
\end{remark}

In Section~\ref{subsec_jcmg} we exploit recursion \eqref{eq.2dARMA} to model network processes for the task of prediction. This will allow us to exploit its polynomial implementation, and thus provide a distributed predictor.

\subsection{Stationarity on Networks}
\label{subsec_stat}

Stationarity is another useful property that extends to graph signals \cite{Girault2015,Perraudin2016,Marques2016} and allows performing graph Wiener filtering/interpolation and spectral estimation. \rev{From the analogies with the concept of stationarity in time signals, graph stationarity requires that the signal first and second order moments are preserved along the graph dimension.} Limiting ourselves to zero-mean processes, one way to define graph wide-sense stationarity (GWSS) is:

\begin{definition}[GWSS]\label{def.gwss} A random graph process $\x$ is wide-sense stationary, if and only if its covariance matrix $\bSigma_x = \Exp[\x\x^\herm]$ is diagonalizable by the GFT basis $\U_\Gr$, \ie $\bSigma_x = \U_\Gr\textnormal{diag}(\p_x)\U_\Gr^{\herm}$. The $N \times 1$ vector $\p_x$ is the GFT of the autocorrelation function of $\x$ and is referred to as the graph power spectral density.
\end{definition}

This condition imposes the invariance of the second order statistics w.r.t. a translation using the shift operator $\L_\Gr$. \rev{That is, as for time signals the second order moment is invariant to temporal translations, here, the invariance of the graph signal statistics should also account for the underlying structure. Moreover, similar to classical temporal stationarity, a useful property of GWSS signals is that $\x$ can be driven by a graph filter applied to a zero-mean white noise.} The graph filter $h(\LAM_\Gr)$ should then satisfy $|h(\LAM_\Gr)|^2 = \text{diag}(\p_x)$.

In analogy with the definition of the JFT, stationarity can be extended to time-varying graph processes \cite{loukas2016stationary}. \rev{This new concept of stationarity requires now that the first and second order moments are jointly  preserved along the graph and temporal dimension.} Limiting ourselves again to zero-mean processes, joint wide-sense stationarity (JWSS) can be defined as follows:

\begin{definition}[JWSS]\label{def.jwss}
A random time-vertex process $\x_{1:T} = \textnormal{vec}(\X_{1:T})$ is called jointly (or time-vertex) wide-sense stationary, if and only if its covariance matrix $\bSigma_{x_{1:T}} = \Exp[\x_{1:T}\x_{1:T}^\herm]$ is diagonalizable by the JFT basis $\U_{\mathcal{J}}$, \ie $\bSigma_{{1:T}} =  \U_{\mathcal{J}}\textnormal{diag}(\p_{{1:T}})\U_{\mathcal{J}}^{\herm}$. The $NT\times 1$ vector $\p_{{1:T}}$ is the JFT of the autocorrelation function of the process $\x_{1:T}$ and is referred to as joint power spectral density (JPSD). 
\end{definition}
%

With this definition, one now assumes simultaneously wide-sense stationarity w.r.t. both the \emph{time} and \emph{vertex} domains. This condition requires now invariance w.r.t. translation using the joint time-vertex shift operator. Further, for a JWSS time-vertex process the covariance is now related to a joint filter $h(\LAM_\Gr,\LAM_{\mathcal{T}})$. Specifically, $\x_{{1:T}}$ can be driven by time-vertex filtering zero-mean white noise, where $h(\LAM_\Gr,\LAM_{\mathcal{T}})$ must now satisfy $|h(\LAM_\Gr,\LAM_{\mathcal{T}})|^2 = \text{diag}(\p_{{1:T}})$.


\section{Modeling of Network Processes}
\label{sec.mod}

This section formulates the problem of modeling the temporal evolution of a JWSS network process. We start with a joint non-causal model and highlight its implementation issues. Then, we follow the classical literature of VARMA recursions \cite{lutkepohl2005new} to propose joint causal models for graph-time series. Finally, the optimal MSE predictor is derived.

\subsection{Joint Non-Causal Models on Graphs}
\label{subsec_jncmg}
A JWSS process $\x_{1:T}$ can be obtained through filtering of a random white process $\eps_{1:T} = \text{vec}(\E_{1:T})$ with a joint time-vertex filter \eqref{eq:def_joint_filtering} (recall Definition~\ref{def.jwss}). Specifically, we can model $\x_{1:T}$ with the \emph{joint non-causal model}
\begin{equation}
\label{eq.jncm}
a(\L_{\mathcal{J}})\x_{1:T} = b(\L_{\mathcal{J}})\eps_{1:T},
\end{equation}
where $\eps_{1:T}$ is the innovation vector, \ie a random vector with zero-mean and identity covariance matrix. The $NT \times NT$ matrices $a(\L_{\mathcal{J}}) = \U_{\mathcal{J}}a(\LAM_\Gr,\LAM_{\mathcal{T}})\U_{\mathcal{J}}^\herm$ and $b(\L_{\mathcal{J}}) = \U_{\mathcal{J}}b(\LAM_\Gr,\LAM_{\mathcal{T}})\U_{\mathcal{J}}^\herm$ represent arbitrary joint filters. From \eqref{eq.jncm} and the statistics of $\eps_{1:T}$, the JPSD of $\x_{1:T}$ is
\begin{equation}
\label{eq.jpsd_xT}
\p_{1:T} = \text{diag}\Bigg(|a(\LAM_\Gr,\LAM_{\mathcal{T}})|^{-2}|b(\LAM_\Gr,\LAM_{\mathcal{T}})|^2\Bigg)
\end{equation}
\rev{for an invertible matrix model $a(\L_{\mathcal{J}})$.}

Despite its generality, model \eqref{eq.jncm} presents several issues for the task of prediction. First, its computational complexity results in a heavy task even for moderate values of $N$ and $T$.
%
%
\rev{Second, \eqref{eq.jncm} is not causal.} This is problematic for prediction where one needs to forecast the future in a timely manner from the process history. To accomplish these challenges, next we introduce \emph{joint causal models}, which ease the computational costs and model the JWSS graph process in a causal manner.

\subsection{Joint Causal Models on Graphs}
\label{subsec_jcmg}

In the following, we introduce two models for forecasting time series on graphs. 

\vskip 1mm
\noindent\textbf{Graph-VARMA model.} Similar to the multivariate case~\cite{lutkepohl2005new}, a joint causal model on graphs has the form
\begin{align}
\x_t = -\sum_{p = 1}^P a_p(\L_\Gr) \, \x_{t-p} +  \sum_{q = 0}^Q b_{q}(\L_\Gr) \, \eps_{t-q},   
\label{eq:joint_causal_model}
\end{align}
where the $N \times N$ matrices $a_p(\L_\Gr) = \U_\Gr a_p(\LAM_\Gr)\U_\Gr^\herm$, $b_0(\L_\Gr) = \I_N$, and  $b_q(\L_\Gr) = \U_\Gr b_q(\LAM_\Gr)\U_\Gr^\herm$ are graph filters and $\eps_t$ is a random vector with zero-mean and covariance matrix $\bSigma_\varepsilon$. 

Following the convention of time series analysis, we call the above model a graph-VARMA (G-VARMA) model. Nevertheless, we also refer to it as a joint causal model because (for sufficiently large $P$ and $Q$) \eqref{eq:joint_causal_model} can describe all causal JWSS processes.

\begin{remark}[G-VARMA complexity]
The per-iteration complexity of the G-VARMA model is $O((P+Q)N^2)$ amounting to $P+Q$ matrix-vector multiplications. 
\end{remark}

{In the sequel, we specialize model \eqref{eq:joint_causal_model} with graph filters $a_p(\L_\Gr)$ and $b_q(\L_\Gr)$ that have a polynomial structure over the graph. With the aim to implement a distributed predictor, we drop the MA part of \eqref{eq:joint_causal_model} and consider graph polynomial-VAR (GP-VAR) models. The rationale beyond the latter choice is that a graph polynomial VARMA model leads in practice to unstable predictors, or in best cases to predictors with larger MSE than GP-VAR. To some degree, this observation is not surprising and goes in line with the conclusions in multivariate modeling \cite[Chapter~11]{lutkepohl2005new}, \cite[Chapter~2]{feng2016signal}. Differently, this behaviour is not consistently observed for model \eqref{eq:joint_causal_model} due to its fitting procedure in the graph Fourier domain (see Proposition~\ref{prop.decoup} in Section~\ref{subsec_cgvarma}).}

\vskip 1mm
\noindent\textbf{Graph polynomial-VAR model.} By exploiting the proposed joint time-vertex filter \eqref{eq.2dARMA}, we write the matrices $a_p(\L_\Gr)$ and $b_{q}(\L_\Gr)$ as polynomials of the graph Laplacian $\L_\Gr$ and get the GP-VAR model\footnote{We denote the approach of \cite{mei2017signal} as restricted graph polynomial-VAR (RGP-VAR) model.}
\begin{equation}
\label{eq.gcp_varma}
\x_t = - \sum_{p = 1}^P\sum_{l = 0}^{L_p}\psi_{l,p}\L_\Gr^l\x_{t-p} + \eps_t
\end{equation}
which is obtained from \eqref{eq:joint_causal_model} by setting $a_p(\L_\Gr) = \sum_{l = 0}^{L_p}\psi_{l,p}\L_\Gr^l$ for $p = 1, \ldots, P$ and $Q = 0$. \rev{Further, note that the order $L_p$ of the polynomials varies based on the time index.}

Model \eqref{eq.gcp_varma} preserves the recursive implementation of \eqref{eq:joint_causal_model}, thus it reduces further the computational complexity. The subsequent remark quantifies this cost.

\begin{remark}[GP-VAR complexity]\label{lemma.dist_complex} \rev{The Laplacian polynomial form of \eqref{eq.gcp_varma} allows it to be implemented distributively in the vertex domain with a complexity similar to \cite{Isufi2016separable}. By setting the maximum order of the Laplacian polynomial as $L_{\textit{max}} = max\{L_1, \ldots, L_p\}$, the per-iteration complexity of the GP-VAR model (13) is $O(L_{\textit{max}}|\mathcal{E}|)$.}
\end{remark}


Since in general $|\Ed| \ll N^2$, the (distributed) per-iteration computational complexity of the GP-VAR model \eqref{eq.gcp_varma} is much smaller than that of the G-VARMA model \eqref{eq:joint_causal_model}. Compared to the non-causal model \eqref{eq.jncm}, if we arrest the recursion after $T$ iterations, the computational complexity of the GP-VAR is $O(L_{\text{max}}|\Ed|T)$, while that of the G-VARMA is $O((P+Q)N^2T)$. Moreover, w.r.t. the RGP-VAR model of \cite{mei2017signal}, \eqref{eq.gcp_varma} allows the polynomial order in the graph domain $L_p$ to differ for each $p$ and not to be restricted as $L_p \le p$. The latter allows capturing more hidden interactions between nodes and, therefore, yields better predictors.
%
%
%
In Section~\ref{subsec_cgpvar}, we will see that this polynomial implementation simplifies the model design, to that of finding scalar coefficients rather than matrices as in \eqref{eq:joint_causal_model}. 



\subsection{Optimal predictor}
\label{subsec_pred}

Let $\X_{1:T}$ be the output of the more general G-VARMA model \eqref{eq:joint_causal_model}. We consider the realizations $\{\x_1, \ldots, \x_{t-1}\}$ are observed, and the goal is to estimate $\x_t$ from these values. Following the conventional approach \cite{lutkepohl2005new}, the value of $\x_t$ is estimated as the conditional expectation
\begin{align}\label{eq.one_stepPred}
\begin{split}
\tilde{\x}_t &= \Exp[\x_{t|t-1}]\triangleq \Exp\bigg[\x_t|\{\x_1, \ldots, \x_{t-1}\}\bigg]\\
&= - \sum_{p = 1}^P a_p(\L_\Gr)\, \x_{t-p} + \sum_{q = 0}^Q b_q(\L_\Gr) \Exp[{\eps_{t-q|t-1}}],
\end{split}
\end{align}
i.e., the expected value of $\x_t$ given the past realizations $\{\x_1, \ldots, \x_{t-1}\}$. Similarly, $\Exp[{\eps_{t-q|t-1}}]$ is the conditional expectation of $\eps_{t-q}$ given $\{\x_1, \ldots, \x_{t-1}\}$. Note that, since the past realizations $\{\x_1, \ldots, \x_{t-1}\}$ follow model \eqref{eq:joint_causal_model}, the knowledge of $\{\x_1, \ldots, \x_{t-1}\}$ allows computing the value of the \emph{specific realizations} for the past innovations $\{\eps_1, \ldots, \eps_{t-1}\}$ deterministically as $\eps_{t-q|t-1} = \x_{t-q|t-1} - \tilde{\x}_{t-q}$ for $q = 1, \ldots, Q$ (recall $b_0(\L_\Gr) = \I_N$). So, we can rewrite \eqref{eq.one_stepPred} as
\begin{align*}
\begin{split}
    &\tilde{\x}_{t} 
    = - \sum_{p = 1}^P a_p(\L_\Gr) \x_{t-p}  + \b_0(\L_\Gr)\Exp[{\eps_{t}}] \\
    &\qquad\qquad\qquad + \sum_{q = 1}^Q b_q(\L_\Gr)\left(\x_{t-q|t-1} - \tilde{\x}_{t-q}\right)
\end{split}
\end{align*} 
\begin{equation}\label{eq:predictor1}
\quad= -\! \sum_{p = 1}^P a_p(\L_\Gr)\, \x_{t-p} \!+\! \sum_{q = 1}^Q b_q(\L_\Gr) \left(\x_{t-q|t-1} \!-\! \tilde{\x}_{t-q}\right).
\end{equation}
where in the last equality we used $\Exp[\eps_t] = \mathbf{0}_N$. We refer to $\tilde{\x}_{t}$ as the \emph{one-step ahead predictor}. The $k-$step ahead predictor can be obtained by repeating the above computation $k$ times.

The one-step ahead predictor for the GP-VAR model \eqref{eq.gcp_varma} can be obtained from \eqref{eq:predictor1} by setting $a_p(\L_\Gr) = \sum_{l = 0}^{L_p}\psi_{l,p}\L_\Gr^l$ for $p = 1, \ldots, P$ and $Q = 0$. In the sequel, we show that \eqref{eq:predictor1} yields the optimal decision from an MSE perspective.

\subsection{Mean Square Error Analysis}
\label{subsec_mse}

We now illustrate the impact of the predictor parameters $a_p(\L_\Gr)$ and $b_q(\L_\Gr)$ on the MSE of the one-step ahead predictor \eqref{eq:predictor1}.

The main result is summarized in the following proposition.

\begin{proposition}
\label{prop:mse_1step}
Let $\x_t$ be the output of the joint causal model \eqref{eq:joint_causal_model} and let $a_p(\L_\Gr)$ and $b_q(\L_\Gr)$ be a given set of model parameters. Then, the MSE of the one-step ahead predictor \eqref{eq:predictor1} is
\begin{equation}
\label{eq:MSE1Step}
\text{MSE} =  \Exp\left[\|b_0(\L_\Gr)\eps_t\|_2^2		\right] = \trace\left(\bSigma_{\varepsilon}	\right),
\end{equation}
which corresponds to the smallest achievable MSE.
\end{proposition}

(The proof is given in the Appendix.)

\section{Model fitting: Parameter Design}
\label{sec.fit}

In this section, we focus on identifying the model parameters, \ie the kernel matrices $a_p(\L_\Gr)$ and $b_q(\L_\Gr)$ for the G-VARMA, or the scalars $\psi_{l,p}$ for the GP-VAR, that fit process $\X_{1:T}$. We start with the more complex G-VARMA case and introduce a GFT-based decoupling approach to ease the computational burden. To further reduce the parameter estimation costs, we introduce a low-rank model which trades prediction accuracy with computational complexity. Then, we consider the GP-VAR predictor and show that the scalar parameters can be estimated from the graph process statistics. The section is concluded with the notion of joint frequency smoothing and its application to parameter design.

\subsection{G-VARMA fitting}
\label{subsec_cgvarma}

Given the realizations $\X_{1:T}$, the canonical way for estimating the kernel matrices $a_p(\L_\Gr)$ and $b_q(\L_\Gr)$ is by solving

\begin{equation}
\label{eq.gcvarma_estim}
\minimize_{a_p(\L_\Gr), b_q(\L_\Gr)} \sum_{\tau = \text{max}\{P,Q\}}^{T-1}\bigg\|\x_\tau - \tilde{\x}_{\tau}[a_p(\L_\Gr), b_q(\L_\Gr)]	\bigg\|_2^2.\\
\end{equation}
{Problem~\eqref{eq.gcvarma_estim} involves a non-linear system of $N \times (T-1- \text{max}\{P,Q\})$ equations with $(P+Q)N$ unknowns whose computational complexity can become prohibitive even for moderate values of $N$ and $T$.
%
%
%
To alleviate these costs, the following proposition establishes a decoupling approach that breaks down problem \eqref{eq.gcvarma_estim} into $N$ uncorrelated (yet not equivalent) well-studied problems with smaller complexity.}

\begin{proposition}[Decoupling]\label{prop.decoup} Consider the joint causal model~\eqref{eq:joint_causal_model}
\begin{align*}
\x_t = -\sum_{p = 1}^P a_p(\L_\Gr) \, \x_{t-p} +  \sum_{q = 0}^Q b_{q}(\L_\Gr) \, \eps_{t-q},   
\end{align*} 
and denote with $\hat{\varepsilon}_{t,n} = [\U_\Gr^{\herm} \eps_{t}]_n$ and $\hat{x}_{t,n} = [\U_\Gr^{\herm} \x_{t}]_n$ the $n$th GFT coefficient of $\eps_{t}$ and $\x_{t}$, respectively. Then, the input-output relation between $\hat{\varepsilon}_{t,n}$ and $\hat{x}_{t,n}$ is given by the ARMA recursion 

\begin{align}
     \hat{x}_{t,n} = - \sum_{p = 1}^P \left[a_p(\LAM_\Gr)\right]_{n,n} \hat{x}_{t-p,n} + \sum_{q = 0}^Q \left[b_q(\LAM_\Gr)\right]_{n,n} \, \hat{\varepsilon}_{t-q,n},
    \label{eq:prop2}
\end{align} 
with $\left[b_0(\LAM_\Gr)\right]_{n,n} = 1~\forall n$ and $n = 1, \ldots, N$.
\end{proposition}

(The proof is given in the Appendix.)

Although simple in its derivation, Proposition \ref{prop.decoup} relates different aspects of the graph and the time series with the model fitting. First, it uses the GFT eigenbasis to formulate \eqref{eq:joint_causal_model} in the graph Laplacian eigenspace\footnote{Recall that for JWSS graph processes the eigenvectors $\U_\Gr$ of $\L_\Gr$ coincide with the eigenvectors of $\bSigma_x$.}. Second, the model fitting is performed in the graph frequency domain. Here, the parameter estimation is split into $N$ uncorrelated problems of a smaller complexity involving $T$ equations and $P+Q$ unknowns. Despite being non-linear problems, the estimation for each graph frequency time series results in fitting a univariate temporal ARMA. So, we can use well-studied methods to solve it, such as the Gauss-Newton approach \cite{wills2008gradient}. 

\begin{remark}[Estimation cost]
In our analysis, the eigendecomposition of the graph Laplacian $\L_\Gr$ is crucial to estimate the G-VARMA parameters. Therefore, the proposed framework suits better small to medium-sized graphs, where the eigenvalue decomposition cost (inherent in joint models), is overshadowed by that of the model estimation. Nevertheless, next, we introduce a low-rank model estimation that reduces the estimation complexity at the expense of fitting accuracy.
\end{remark}

\noindent\textbf{Low-rank models.}  To further reduce the model estimation cost, we can consider estimating $a_p(\L_\Gr)$ and $b_q(\L_\Gr)$ only from a subset of graph frequencies, say $K \le N$. The latter incurs considerable savings in terms of estimation complexity since $(i)$ there is no need to compute the full eigendecomposition of $\L_\Gr$, but only the eigenvectors relative to the chosen $K$ graph frequencies; and $(ii)$ we only need to fit $K \le N$ temporal ARMA time series to the data. For this purpose, we use the following definition of low-rank matrix approximation.

\begin{definition}[Low-rank approximation] Let $\Scal$ be an index set of cardinality $K = |\Scal|$, with indicator matrix $\D_\Scal$. Let also $\U$ be a unitary (rotation) matrix. The $\{\U, \Scal\}$ low-rank approximation of a matrix $\X_{1:T}$ is
\begin{equation}
\tilde{\X}_{\U,\Scal} = \U\D_\Scal\U^\herm\X_{1:T}.
\end{equation}
\end{definition}
For prediction, which is an online task, the above low-rank approximation should be done only w.r.t. the graph dimension. 
The following theorem provides a constructive way to design the set $\Scal$ that achieves the \emph{optimal low-rank} approximation for a JWSS process.

\begin{theorem}\label{theo.low_rank}
Let $\X_{1:T}$ be a zero-mean JWSS process. The optimal K-rank approximation of $\X_{1:T}$ is given by
\begin{equation*}
\label{eq.theo_LR}
\{\U_\Gr,\Scal^\star	\} = \argmin_{\U, \Scal}\Exp\left[	\left\|	\X-\tilde{\X}_{\U,\Scal}	\right\|^2_F	\right]\quad \text{s.t.}\quad |\Scal| = K,
\end{equation*}
where $\Scal^\star$ contains the indices of the $K$ largest diagonal elements of $\U_\Gr^\herm\bSigma_{x}\U_\Gr$.
\end{theorem}

(The proof is given in the Appendix.)

This result suggests that the best rank$-K$ approximation of $\X_{1:T}$ is obtained by rotating the JWSS process by the graph Laplacian eigenvectors. This finding is beneficial since the first step in the model estimation consists of decoupling the time series using the GFT. It further allows us to claim the following:


\begin{corollary}\label{cor_transf} Let $\U_{\Gr,K}$ denote the $N \times K$ matrix containing the $K$ columns of $\U_\Gr$ relative to the $K$ {highest} eigenvalues of $\bSigma_x$. Then, $\U_{\Gr,K}^\herm\bSigma_{x}\U_{\Gr,K}$ leads to the same result as selecting the {highest} K diagonal elements of $\U_\Gr^\herm\bSigma_{x}\U_\Gr$.
\end{corollary}

An important outcome of the last two results is that the full eigendecomposition of $\L_\Gr$ is not needed. Indeed, only the eigenvectors relative to the $K$ largest eigenvalues of $\bSigma_{x}$ are needed (recall that $\U_\Gr$ jointly diagonalizes both $\L_\Gr$ and $\bSigma_{x}$), which can be obtained with a lower cost.
Therefore, the low-rank representation can be combined with the model estimation, by only modeling the time series for the graph frequencies in $\Scal$. In this way, we attain a reduction in the model estimation cost of $N/K$ allowing the G-VARMA to cope well with largely-sized graphs.
Later in Section~\ref{subsec_lowRank}, we will see that real data enjoy this low-rank representation with $K \ll N$.


\begin{figure*}[!b]
\hrulefill
\begin{align}
\begin{split}
&\underset{\{\psi_{l,p}\}}{\text{minimize}}\quad \trace\left(\R_x(0)	 +\sum_{p=1}^P\bPsi_p\R_x(p) +\sum_{p=1}^P\R_x(p)\bPsi_p^\herm + \sum_{p_1 = 1}^P\sum_{p_2 = 1}^P\bPsi_{p_1}\R_x(p_2-p_1)\bPsi_{p_2}^\herm\right)\\
&\text{subject to} \quad \bPsi_p = \sum_{l=0}^{L_p}\psi_{l,p}\L_\Gr^l.
\end{split}
\label{eq.gc_varprob}
\end{align}
\end{figure*}

\subsection{GP-VAR fitting}
\label{subsec_cgpvar}

The parameter estimation for the GP-VAR model \eqref{eq.gcp_varma} consists of estimating the scalars $\psi_{l,p}$. Since this problem has a lower complexity, we find the $\psi_{l,p}$ coefficients by directly minimizing the prediction MSE
\begin{equation}
\label{eq.gcp_MSE1}
\text{MSE} = \Exp\Bigg[\Bigg\|\x_t + \sum_{p=1}^P\bPsi_p\x_{t-p} 	\Bigg\|_2^2\Bigg],
\end{equation}
with $\bPsi_p = \sum_{l=0}^{L_p}\psi_{l,p}\L_\Gr^l$. From $\|\x\|_2^2 = \trace(\x\x^\herm)$ and the linearity of the trace and the expectation, \eqref{eq.gcp_MSE1} becomes
\begin{align}
\label{eq.gcp_MSE2}
\begin{split}
\text{MSE} &= \trace\left(\sum_{p=1}^P\bPsi_p\Exp\left[ \x_{t-p}\x_t^\herm	\right]	\right) + \trace\left(\sum_{p=1}^P\Exp\left[\x_t\x_{t-p}^\herm\right]\bPsi_p^\herm\right)\\
&\hskip-.6cm+\trace\left(\Exp\left[\x_t\x_t^\herm	\right]	\right) + \trace\left(\sum_{p_1 = 1}^P\sum_{p_2 = 1}^P\bPsi_{p_1}\Exp\left[\x_{t-p_1}\x_{t-p_2}^\herm\right]\bPsi_{p_2}^\herm\right)\!. 
\end{split}
\end{align}
Then, since $ \R_x(i) = \Exp[\x_t\x_{t-i}]$ is the autocorrelation of the process at lag $i$, the GP-VAR coefficients can be found by solving the convex problem \eqref{eq.gc_varprob}.

%

From \eqref{eq.gcp_MSE1}-\eqref{eq.gc_varprob}, the graph topology imposes a structure on the model coefficients. In fact, this, somehow small, modification renders the model parameter estimation a computationally easier task compared to the classical VAR model. Thus, more robust to cope with larger multivariate dimensions. The GP-VAR model inherits several benefits from the classical literature of multivariate VAR analysis \cite{lutkepohl2005new}. In \eqref{eq.gc_varprob}, we rely on the autocorrelation matrix to estimate the coefficients, which must be estimated from the training data $\X_{1:T}$, \ie $\R_x(i) = (T-i)^{-1}\sum_{\tau=0}^{T-i}\x_\tau\x_{\tau+i}^\herm$. Differently from \cite{di2016adaptive, di2017adaptive, Qiu2017, romero2016kernel} that do not exploit stationary assumptions, in the GP-VAR model we can incorporate well-established techniques for estimating $\R_x(i)$, such as the shrinkage estimators \cite{ledoit2004well} and the random matrix theory-based estimator \cite{tulino2004random} in low-samples regime, or the Kullback-Leiber divergence \cite{sun2014regularized} and Tyler's estimator \cite{tyler1987distribution} to deal with heavy tail issues in non-Gaussian scenarios.

\begin{remark}[Yule-Walker estimation] An alternative to \eqref{eq.gc_varprob} for estimating $\psi_{l,p}$ is to follow a Yule-Walker approach \cite{Hayes2009}. That is, minimize in a least squares sense the function $f(\psi_{l,p}) = \R_x(0) + \sum_{p = 1}^P\bPsi_p\R_x(p)$ with $\bPsi_p = \sum_{l=0}^{L_p}\psi_{l,p}\L_\Gr^l$. Besides being a computationally lighter problem to solve, the Yule-Walker strategy can be combined with the approach in \cite{chepuri2017graph} and, thus, estimate the $\psi_{l,p}$ from subsampled measurements.
\end{remark}
%
%

\subsection{Spectral Smoothing}
\label{subsec_specSmooth}

We can further improve the model fitting by using the spectral smoothing heuristic \cite[Section IV. B]{loukas2016stationary}. This involves the convolution of the JPSD with a smoothly decaying windowed function in the spectral domain. Specifically, if $\P_{1:T} = |\JFT{\X_{1:T}}|^2$ is the JPSD in matrix form, we fit the models to the process having a final JPSD
\begin{align}\label{eq.spec_smooth}
	[\tilde{\P}_{1:T}]_{\lambda, e^{-\j\omega}} \!=\!\! \sum_{n,t} g_{\mathcal{G}}(\lambda \!-\! \lambda_n)^2\! g_{\mathcal{T}}(e^{-\j\omega} \!-\! e^{-\j\omega_t})^2 \,\! [\P_{1:T}]_{\lambda_n, e^{-\j\omega_t}},
\end{align}
where $g_{\mathcal{G}}$ and $g_{\mathcal{T}}$ are smoothly decaying functions in the graph and temporal and frequency domains, respectively. 
$g_\Gr$ and $g_{\mathcal{T}}$ are scaled Gaussian functions centered at zero and normalized to sum to one. The Gaussian width is a free parameter that controls the estimator's bias-variance trade-off~\cite{loukas2016stationary}.

The idea of spectral smoothing is common in time series analysis. In fact, for stationary time series, spectral smoothing (i.e., the convolution with $g_{\mathcal{T}}$ along each node) is a technique encountered in the literature (see \cite{bartlett1950periodogram,welch1967use}).
For stationary graph signals, spectral smoothing w.r.t. the graph Fourier domain (i.e., the convolution with $g_{\mathcal{G}}$ along each $t$) is used to better estimate stationary graph signals~\cite{Perraudin2016, girault2017towards}.
This heuristic also relates to the spectral leakage phenomenon in eigenvalue estimation~\cite{loukas2017close}, exploiting the locality of spectral leaking to better estimate the eigenvalues of the covariance matrix from few samples.

With respect to this work, \eqref{eq.spec_smooth} relates the joint stationarity of the signal with the challenge of estimating the model parameters in a low sample regime. This heuristic exploits the fact that the signal JPSD is likely to have similar values in adjacent frequencies and that abrupt changes are more likely to occur due to noise and finite sample estimation inaccuracies. Therefore, by exploiting the latter, a model fitting to data with a JPSD as in \eqref{eq.spec_smooth} will often lead to a lower estimation error.

\section{Tracking on a subset of nodes}
\label{sec.subset}
\begin{figure}
\centering
\includegraphics[width=0.8\columnwidth]{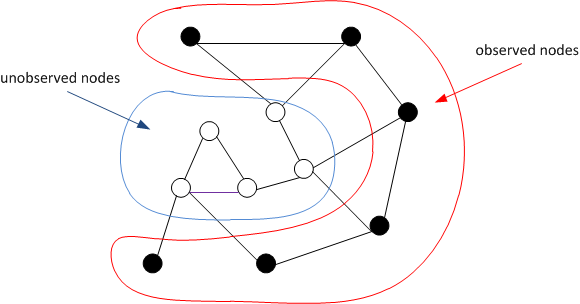}
\caption{Illustration of the sampling set and its complementary set. The black nodes are observed at time $t$, while the white nodes are not accessed. At previous time instants, all nodes are observed.}
\label{fig.ill_graph}
\end{figure}

This section looks at estimating $\x_t$ by using both the historical data and measurements at time $t$ collected on a subset $\Scal_t \subseteq \Vx$ of nodes. In what follows, we consider the G-VARMA model to address this task. At the end of this section, we remark how the tracking can be addressed by other models.

A possible application involving this scenario comprises a survey outcome in a social network. Here, it is likely that a (possibly significant) number of users do not take part in the survey after a certain time instant.
%
By using the developed predictors and the observations on $\Scal_t$, we are interested to estimate the survey outcome at time $t$ on another set $\bar{\Scal}_t$. 
The set $\bar{\Scal}_t$ might then be: $(i)$ $\bar{\Scal}_t = \Vx$, i.e., improve the estimation on all nodes; $(ii)$ $\bar{\Scal}_t = \Vx\backslash\Scal_t$, i.e., the complementary set of $\Scal_t$; or $(iii)$ a combination on the two. Fig.~\ref{fig.ill_graph} illustrates one such example.

Let $\D_t \in \{0,1\}^{|\Scal_t|\times N}$ be a binary matrix containing the non-zero rows of $\D_{\Scal_t} = \text{diag}(\mathbf{1}_{\Scal_t})$. Note that by construction $\D_t\D_t^\transp = \I_{|\Scal_t|}$ and $\D_t^\transp\D_t = \D_{\Scal_t}$. Similarly, let $\bar{\D}_t$ be the $|\bar{\Scal}_t| \times N$ binary matrix relative to the set $\bar{\Scal}_t$.
%
%
%
%
Let also $\z_t = \D_t(\x_t + \w_t)$ be the $|\Scal_t| \times 1$ vector of measurements at time $t$ with $\w_t$ the zero-mean measurement noise with covariance matrix $\bSigma_w$ independent from $\boldsymbol{\varepsilon}_t$. Then, for $\x_t$ following the G-VARMA model \eqref{eq:joint_causal_model}, we have
\begin{equation}
\z_t = \D_t \eps_t + \D_t\tilde{\x}_{t} + \D_t\w_t,
\end{equation}
with $\tilde{\x}_{t}$ from \eqref{eq:predictor1}. The linear minimum mean square estimate (LMMSE) for $\eps_t$ is
\begin{align}\label{eq.lmmse1}
\tilde{\eps}_t &=  \bSigma_{\varepsilon}\D_t^\transp \H_t^{-1}(\z_t - \D_t\tilde{\x}_{t}),
\end{align}
where $\H_t = \D_t\left(\bSigma_{\varepsilon} + \bSigma_w\right)\D_t^\transp$ \cite{kay1993fundamentals}. Substituting $\tilde{\eps}_t$ into the signal evolution over the set $\bar{\Scal}_t$, we get
\begin{align}\label{eq.lmmse_fin}
	\tilde{\x}_{\bar{\Scal},t} = \bar{\D}_t \bigg(\tilde{\x}_{t} + \bSigma_{\varepsilon}\D_t^\transp \H_t^{-1}(\z_t - \D_t\tilde{\x}_{t}) \bigg). 
\end{align}
This is the estimate of $\x_t$ on $\bar{\Scal}_t$ which considers the G-VARMA prediction $\tilde{\x}_t$ and the measurements collected over $\Scal_t$. Estimator \eqref{eq.lmmse_fin} considers, in addition to its past realizations, i.e., $ \bar{\D}_t\tilde{\x}_{t}$, an inverse filtering on the observed vertices and then a spreading onto the set $\bar{\Scal}_t$.

Following the KF convention \cite{simon2006optimal}, $\tilde{\x}_{\bar{\Scal},t}$ in \eqref{eq.lmmse_fin} is the \emph{a posteriori} estimate of $\x_t$ on the set of interest $\bar{\Scal}_t$ and $\tilde{\x}_{t}$ is the \emph{a priori} estimate of $\x_t$. The latter allows then to easily quantify in closed form the MSE of estimator \eqref{eq.lmmse_fin}. In fact, given the \emph{a priori} error covariance matrix over $\mathcal{V}$, $\P_t^- = \bSigma_\varepsilon$, and the \emph{a posteriori} error covariance matrix over ${\bar{\Scal}_t}$
\begin{equation}
\label{eq.pos_cov_mat}
\P_{\bar{\Scal},t}^+ = \bar{\D}_t(\I_N - \K_t\D_t)\P_t^-\bar{\D}_t^\transp,
\end{equation}
with Kalman gain matrix
\begin{equation}\label{eq.KalGain}
\K_t = \P_t^-\D_t^\transp\H_t^{-1},
\end{equation}
the MSE of \eqref{eq.lmmse_fin} is
\begin{align}\label{eq.MSE_subset}
\begin{split}
\textit{MSE} &= \Exp\bigg[\|\x_{\bar{\Scal},t} - \bar{\D}_t\x_t	\|_2^2	\bigg]\\
&=\trace\bigg(\bar{\D}_t(\I_N - \bSigma_\varepsilon\D_t^\transp\H_t^{-1}\D_t) \bSigma_\varepsilon\bar{\D}_t^\transp	\bigg),
\end{split}
\end{align}
with $\H_t$ defined in \eqref{eq.lmmse1}.

The MSE \eqref{eq.MSE_subset} shows the dependence of the reconstruction performance on the sampling set (through $\D_t$) and the G-VARMA model (through the residual error $\bSigma_\varepsilon$). To select the sampling set $\Scal_t$, we can exploit the above KF reformulation of \eqref{eq.lmmse_fin} and use the sparse-sensing sampling strategies \cite{chepuri2016sparse} adopted in \cite{isufi2017observing} for tracking diffusion processes on graphs. However, as found in the latter work, the MSE would improve by $1$dB-$2$dB w.r.t. a random uniform sampling. While the latter can be used (at the expense of a higher complexity), in this work we show the potential of the proposed tracker by building $\Scal_t$ uniformly at random.

\begin{remark}[Other models] As the KF interpretation highlighted, estimator \eqref{eq.lmmse_fin} uses $\P_t^- = \bSigma_\varepsilon$ of G-VARMA to connect the prediction step with the tracking from $\z_t$. Therefore, if we are interested to use another predictive model, the shown KF reformulation can be adopted to incorporate the a priori predictor of $\x_t$ into the measurements $\z_t$. For the GP-VAR, for instance, the $\P_t^-$ can be derived from \eqref{eq.gcp_MSE2}.
%
\end{remark}


\section{Numerical Results}
\label{sec.numRes}

This section tests the proposed models in four different scenarios: the Molene weather data set, the NOAA U. S. temperature data set, a walking dog dynamic mesh, and a simulated susceptible-infected (SI) epidemic spreading over a flight network.

{The predictors in Section~\ref{subsec_jcmg} are compared with: $(i)$ the disjoint univariate ARMA that predicts each time series per node; $(ii)$ the standard VAR \cite{lutkepohl2005new}; $(iii)$ the RGP-VAR from \cite{mei2017signal}; $(iv)$ the LMS on graphs algorithm from \cite{di2016adaptive}; and $(v)$ the RLS on graphs algorithm from \cite{di2017adaptive}. For the LMS and RLS on graphs algorithms, we considered samples collected on all nodes. The tracking performance of the approach introduced in Section~\ref{sec.subset} is compared with the Wiener inpainting \cite{Perraudin2016} (applied at each time instant) and the kernel Kalman filter (KKF) \cite{romero2016kernel}. For the KKF, we considered subsampled measurements only at time $t$.}

\begin{figure}[t]
\centering
\includegraphics[width=.7\columnwidth]{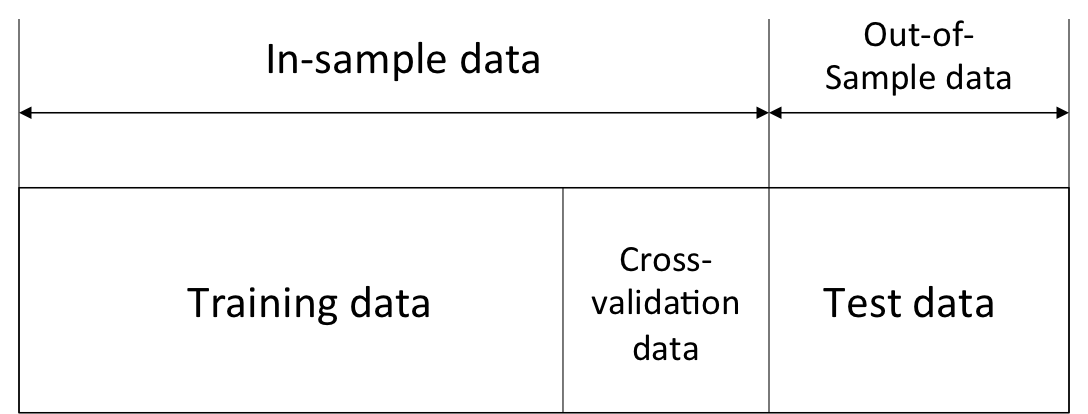}
\caption{Model fitting process \cite{friedman2001elements}.}
\label{fig.fit_proc}
\end{figure}

\rev{For all models, we find the respective parameters by cross-validation, as discussed in \cite{friedman2001elements}.}
With reference to Fig.~\ref{fig.fit_proc}, the time series is split along the temporal dimension into two parts: the in-sample data and the out-of-sample (or testing) data.
The in-sample data are further split into a training and a validation set. The estimation procedure starts with first fitting each model with different model parameters (\eg different values for $P$ and $Q$ in the G-VARMA predictor) to the training data and measures their performance on the validation set. Then, the selected parameters are the one that yield the lowest root normalized MSE (rNMSE) defined as
\begin{equation}\label{eq.av_rNMSE}
\text{rNMSE} = \sqrt{\frac{\sum_{t = 1}^\tau\|\tilde{\boldsymbol{\theta}}_t - \boldsymbol{\theta}_t	\|_2^2}{\sum_{t = 1}^\tau\|\boldsymbol{\theta}_t	\|_2^2}},
\end{equation}
for some unknowns $\boldsymbol{\theta}_1, \ldots, \boldsymbol{\theta}_\tau$ and respective estimates $\tilde{\boldsymbol{\theta}}_1, \ldots, \tilde{\boldsymbol{\theta}}_\tau$. Finally, the model is refitted with the selected parameters to the entire in-sample data and tested on the out-of-sample data. 

We indicate with $\sigma_{\text{g}}$ the Gaussian width used to perform the spectral smoothing along the graph dimension [cf. \eqref{eq.spec_smooth}] and with $\gamma$ the regularization parameter that determines a bias-variance trade-off in finding the ARMA coefficients along the temporal dimension\footnote{This is a regularization parameter used by the Matlab function armax.m that penalizes the $\ell_2$ norm of the ARMA coefficients.} \cite{ljung1998system,wills2005gradient}. In the simulations, we made use of the GSP toolbox \cite{perraudin2014gspbox}.

%

\begin{figure*}[!t]
\centering
\begin{subfigure}[h]{0.255\linewidth}
\includegraphics[width=\linewidth,trim={1.5cm 1.5cm 1.2cm 1cm},clip]{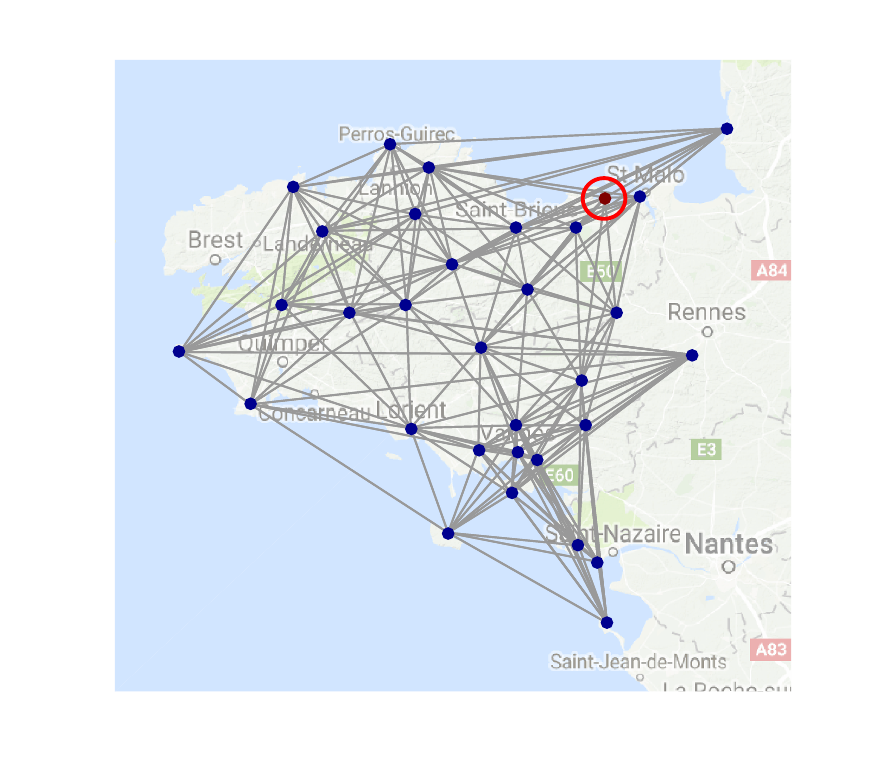}
\caption{}
\end{subfigure}
\hfill
\begin{subfigure}[h]{0.33\linewidth}
\includegraphics[width=\linewidth,trim={0.85cm 0.35cm 0.85cm 0.35cm},clip]{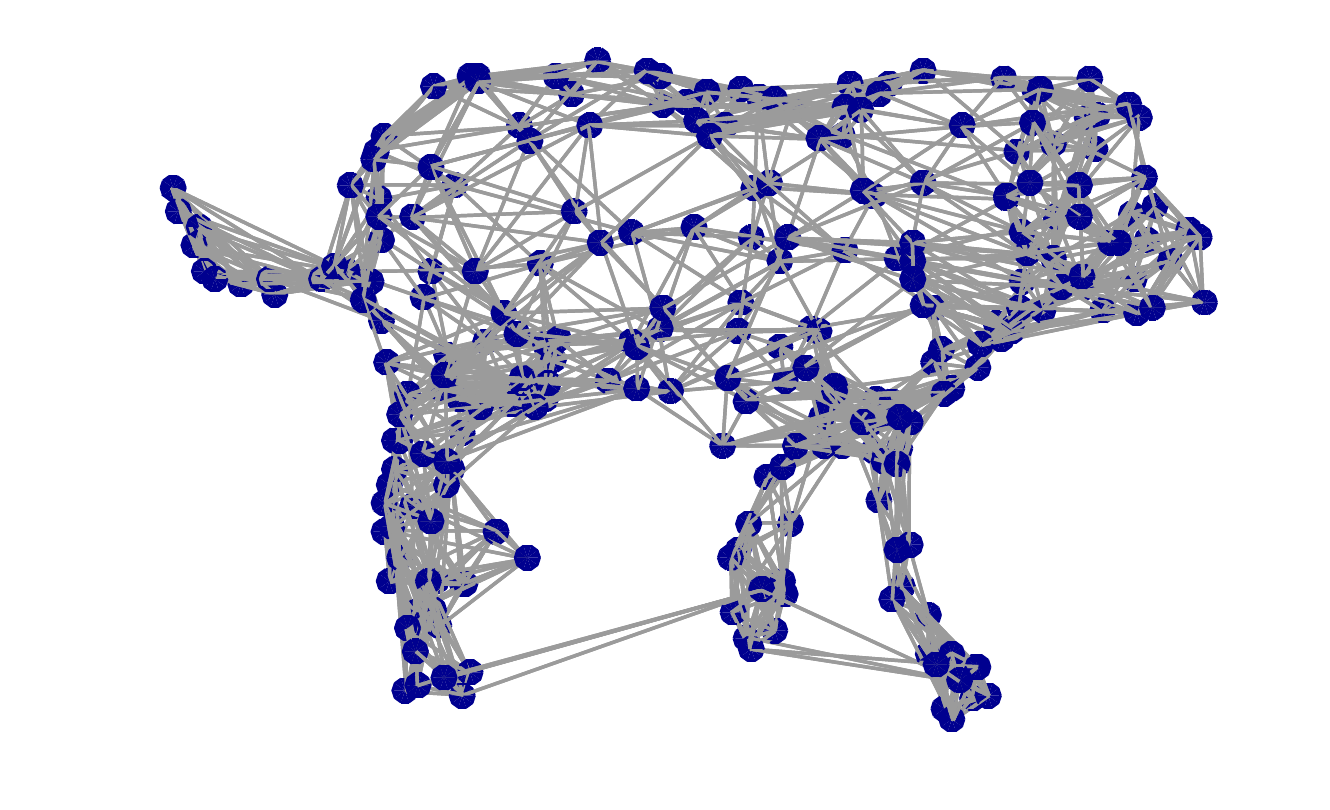}
\caption{}
\end{subfigure}
\hfill
\begin{subfigure}[h]{0.33\linewidth}
\includegraphics[width=\linewidth,trim={.8cm 2cm .8cm 1cm},clip]{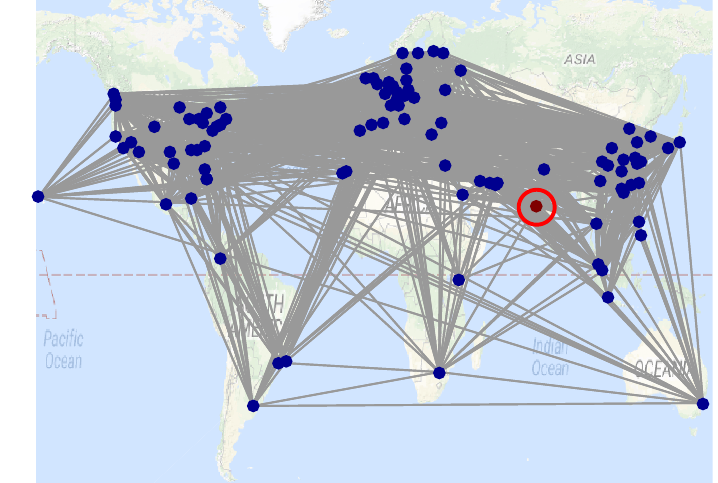}
\caption{}
\end{subfigure}%
\caption{Graph topologies for the considered scenarios. (a) Molene graph. The vertex in red is the one selected to illustrate the prediction performance. \rev{(b) Dog mesh graph. In building the graph, we considered unknown the manifold structure. Thus, we can see edges connecting nodes in different legs that are close by each-other (in the time step on which the graph was constructed).} (c) Airline connectivity graph. \rev{The nodes in blue are in susceptible (S) state, while the random picked node in red (Mumbai) is in the infected (I) state.}}
\label{fig.2Scen_graphs}
\end{figure*}

\subsection{Considered scenarios and data pre-processing}

In the following, we discuss the four datasets used in our evaluation and provide details for graph construction and preprocessing.

\vskip1mm
\noindent\textbf{Molene data set.} The Molene data set contains hourly temperature measurements of $N = 32$ weather stations near Brest (France) for $T = 744$ hours. We consider a geometric graph (illustrated in Fig.~\ref{fig.2Scen_graphs} (a)) \rev{built from the node coordinates using the default nearest neighbor approach of \cite{perraudin2014gspbox}. This graph follows a $10-$NN construction and the weight of the edge $(i,j)$ follows the Gaussian kernel 
\begin{equation*}
[\W_\Gr]_{ij} = \text{exp}\big(-\text{dist}(i,j)/\overline{\text{dist}}	\big)
\end{equation*}
where $\text{dist}(i,j)$ is the Euclidean distance between stations $i$ and $j$ and $\overline{\text{dist}}$ is the average Euclidean distance of all sensors.
}The red station is used later on to illustrate the temperature prediction.
The Laplacian matrix $\L_\Gr$ is normalized to have a unitary spectral norm. Before the fitting process, the in-sample mean is subtracted from the raw data.

\vskip1mm
\noindent\rev{\textbf{NOAA data set.} This is another temperature data set and comprises of hourly temperature recordings at $N = 109$ stations across the United States in 2010 \cite{nooaDataset} for a total of $T = 8759$ hours. This data set has been used by four graph-based alternative techniques that we compare with, respectively the RGP-VAR \cite{mei2017signal}, the LMS and RLS on graphs \cite{di2017adaptive}, and by the KKF \cite{romero2016kernel}. We use the same graph structure as \cite{di2017adaptive} and \cite{romero2016kernel}, which is built following the approach of \cite{mei2017signal} and relies on the $7-$NN geographical distances. The combinatorial Laplacian ${\L}_{\mathcal{G}}$ is used to represent the graph connectivity and the in-sample mean is subtracted from the raw data before preprocessing.
}
\vskip1mm
\noindent\textbf{Walking dog mesh.} We start from the 3D coordinates of a mesh depicting a walking dog~\cite{gall2009motion}. We aim at predicting the average point position along the three coordinates. The mesh has $N = 251$ points (nodes) over $T = 59$ time steps. \rev{The graph follows a $10$-NN construction built from the coordinates at $t = 1$ and is illustrated in Fig.~\ref{fig.2Scen_graphs} (b). This mesh is sparser than the one in  \cite{grassi2017time} (where this data set is used to remove noise) to reduce the computational time.} As a preprocessing step, we again subtract the in-sample mean from the data and, as normalization, we divide by the largest absolute value.

\vskip1mm
\noindent\textbf{SI epidemic diffusion.} This scenario considers an epidemic diffusion following the susceptible-infected model \cite{nowzari2016analysis} over $N = 100$ international airports. The graph, depicted in Fig.~\ref{fig.2Scen_graphs} (c), captures the airline connections with edge weights $[\W_\Gr]_{i,j} = [\W_\Gr]_{j,i} = 1$ if there is a flight connection between two airports and zero otherwise. The graph has $3565$ edges and an average degree of $35.4$. $\L_\Gr$ is normalized to have a unitary spectral norm and the in-sample mean is subtracted from the data.

The time-varying process of interest is the daily evolution of the infection chance of each node (airport). We select an infection rate of $10^{-3}$ and consider a small fixed population occupying each airport (60 targets). 
The infected targets are assumed to return to the susceptible state after 12 days and the overall epidemic diffusion is analyzed for $T = 122$ days. \rev{In the initial state, all targets at the blue nodes are in the susceptible-state (S-state), while all targets at the red node are in the infected-state (I-state).}

\begin{figure*}[!t]
\centering
\fbox{\begin{minipage}{\textwidth}%
    \begin{subfigure}{.5\columnwidth}
\centering
\includegraphics[width=\columnwidth,trim={.3cm 0 .7cm 0},clip]{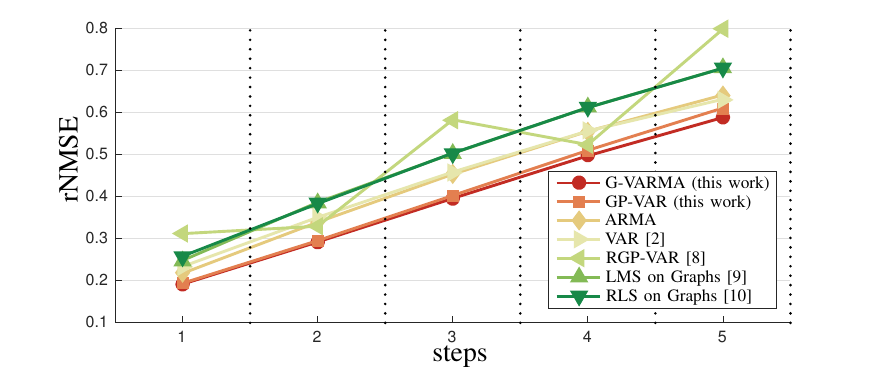}
\caption{rNMSE versus the prediction steps for the different methods. The used parameters are: [G-VARMA \eqref{eq:joint_causal_model} : $P = 3$, $Q = 0$, $\sigma_{\text{g}} = 0$, $\gamma = 0$], [GP-VAR: $P = 2$, $L_p = L = 4$, $\sigma_{\text{g}} = 0$], [ARMA: $P = 4$, $Q = 0$, $\gamma = 0$], [VAR \cite{lutkepohl2005new}: $P = 2$], [RGP-VAR \cite{mei2017signal}: $P = 3$], [LMS on Graphs  \cite{di2016adaptive}: B.width $= 16$, $\mu_{\text{LMS}} = 1.2$], [RLS on Graphs \cite{di2017adaptive}: B.width $=16$, $\beta_{\text{RLS}} = 0.05$].}
\label{fig.mol_pred_pow}
    \end{subfigure}
\hfil
    \begin{subfigure}{.4\columnwidth}
\centering
\includegraphics[width=\columnwidth,trim={.55cm 0 .7cm .35cm},clip]{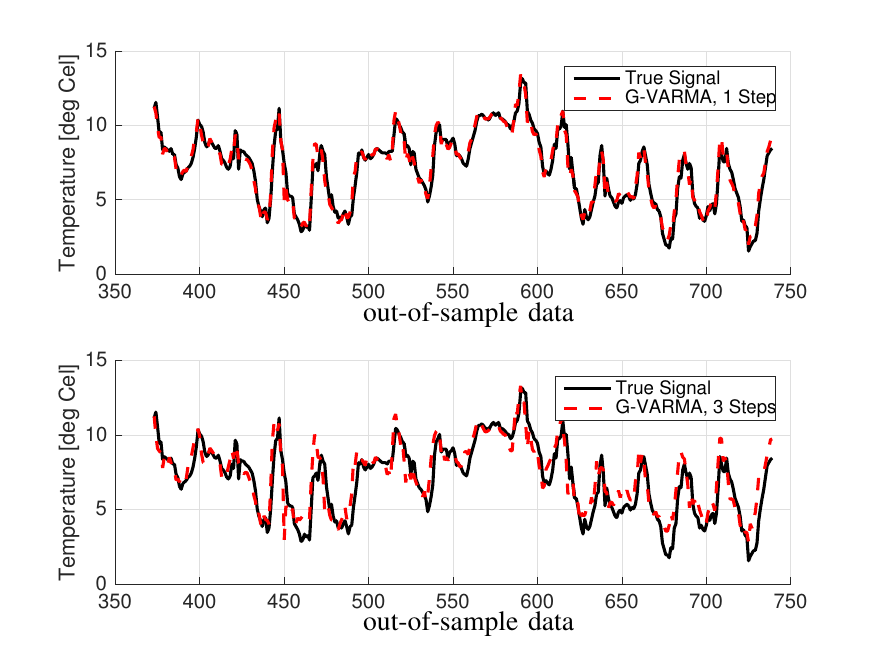}\vskip-2mm
\caption{True temperature in degrees Celsius and the G-VARMA one- and three-steps ahead predictions at the red sensor in Fig.~\ref{fig.2Scen_graphs} (a).}
\label{fig.mol_true_sig}
    \end{subfigure}\\
\vskip2.5mm
    \begin{subfigure}{.5\columnwidth}
    \centering
\includegraphics[width=\columnwidth,trim={.3cm 0 .7cm 0},clip]{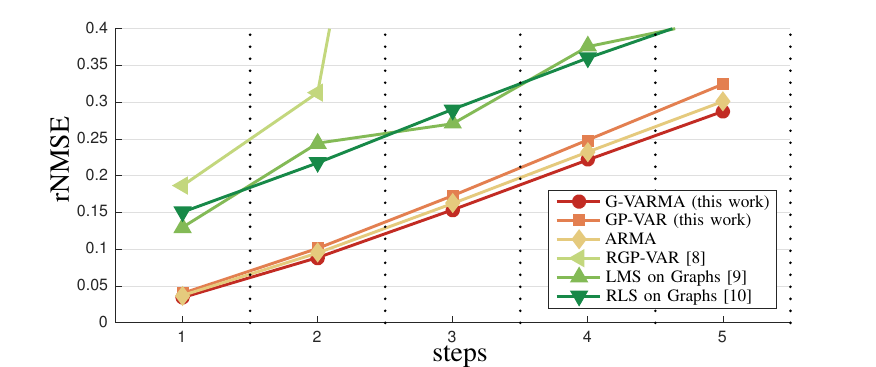}
\caption{rNMSE versus the prediction steps for the different methods. The used parameters are: [G-VARMA \eqref{eq:joint_causal_model} : $P = 3$, $Q = 0$, $\sigma_{\text{g}} = 0$, $\gamma = 0$], [GP-VAR: $P = 3$, $L_p = L = 3$, $\sigma_{\text{g}} = 0$], [ARMA: $P = 3$, $Q = 0$, $\gamma = 0$], [RGP-VAR \cite{mei2017signal}: $P = 2$], [LMS on Graphs  \cite{di2016adaptive}: B.width $= 40$, $\mu_{\text{LMS}} = 1.5$], [RLS on Graphs \cite{di2017adaptive}: B.width $=40$, $\beta_{\text{RLS}} = 0.5$]. A zoomed in version of one one-step ahead prediction is provided in Fig.~\ref{fig.noaaZoom} in the Appendix.}
\label{fig.noaa_pred_pow}
    \end{subfigure}
\hfil
    \begin{subfigure}{.4\columnwidth}
\centering
\includegraphics[width=\columnwidth,trim={.45cm 0 .7cm .35cm},clip]{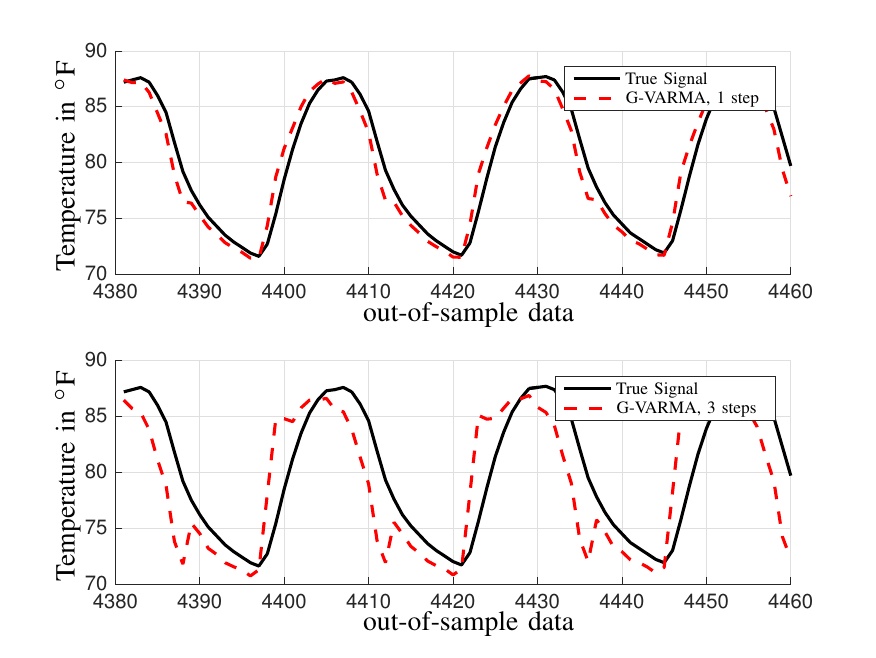}\vskip-2mm
\caption{True temperature in degrees Fahrenheit and the G-VARMA one- and three-steps ahead predictions for the same node visualized \cite{di2017adaptive}.}
\label{fig.noa_true_sig}
    \end{subfigure}
    \end{minipage}}
    \caption{Comparison of the predictive power of the different algorithms in the weather datasets. (a)-(b) Molene data set. (c)-(d) NOAA data set. The results in plots (b) and (d) are w.r.t. the out-of-sample data.}
    \label{fig.motherPred}
    \end{figure*}

\begin{figure*}[!t]
\centering
\fbox{\begin{minipage}{\textwidth}%
    \vskip2.5mm
    \begin{subfigure}{.5\columnwidth}
    \centering
\includegraphics[width=\columnwidth,trim={.3cm 0 .7cm 0},clip]{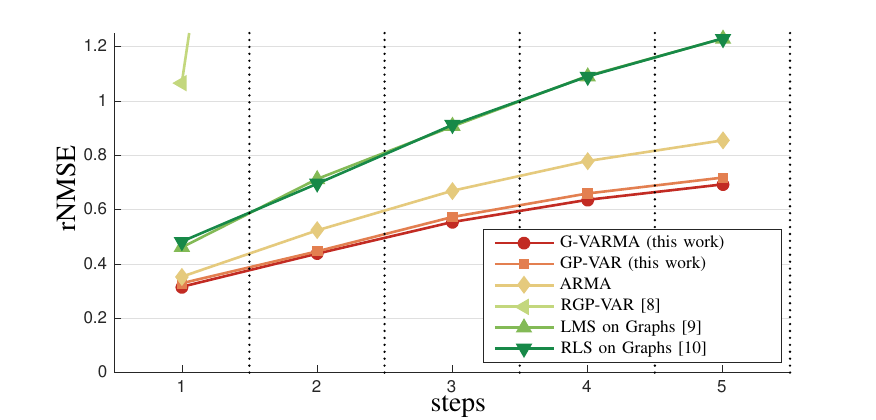}
\caption{rNMSE versus the prediction steps for the different methods. The used parameters are: [G-VARMA \eqref{eq:joint_causal_model} : $P = 3$, $Q = 0$, $\sigma_{\text{g}} = 0$, $\gamma = 0$], [GP-VAR: $P = 4$, $L_p = L = 3$, $\sigma_{\text{g}} = 0.25$], [ARMA: $P = 2$, $Q = 0$, $\gamma = 0$], [RGP-VAR \cite{mei2017signal}: $P = 3$], [LMS on Graphs  \cite{di2016adaptive}: B.width $= 16$, $\mu_{\text{LMS}} = 1.35$], [RLS on Graphs \cite{di2017adaptive}: B.width $=16$, $\beta_{\text{RLS}} = 0.05$].}
\label{fig.dog_pred_pow}
    \end{subfigure}
\hfil
    \begin{subfigure}{.4\columnwidth}
\centering
\includegraphics[width=\columnwidth,trim={.45cm 0 .7cm .35cm},clip]{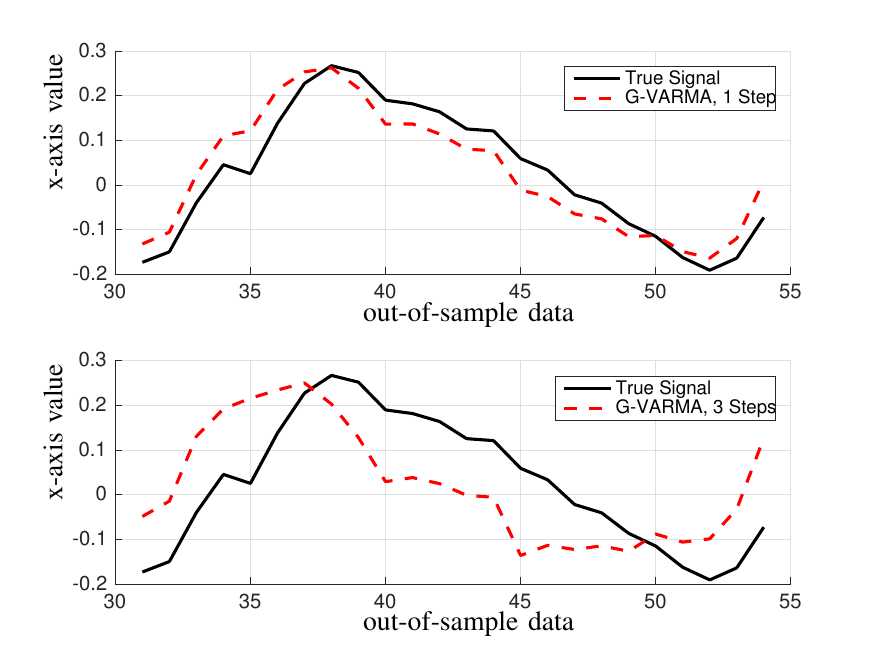}\vskip-2mm
\caption{True x-axis value and the G-VARMA one- and three-steps ahead predictions at the the node that has the largest variability in the true signal.}
\label{fig.dog_true_sig}
    \end{subfigure}\\
\vskip2.5mm
    \begin{subfigure}{.5\columnwidth}
\centering
\includegraphics[width=\columnwidth,trim={.3cm 0 .7cm 0},clip]{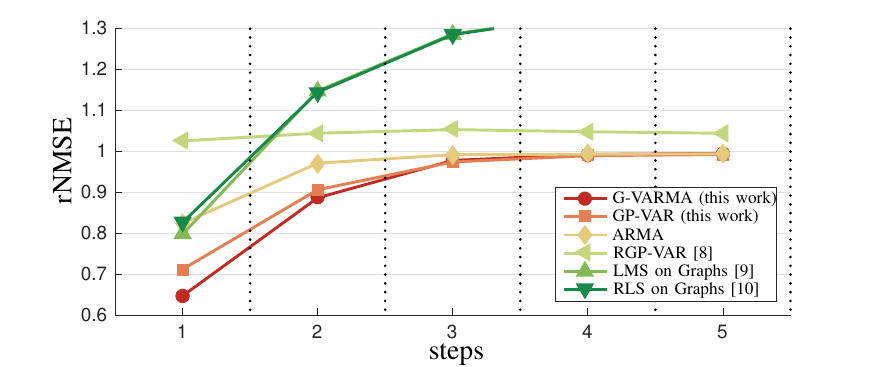}
\caption{rNMSE versus the prediction steps for the different methods. The used parameters are: [G-VARMA \eqref{eq:joint_causal_model} : $P = 4$, $Q = 0$, $\sigma_{\text{g}} = 0$, $\gamma = 0$], [GP-VAR: $P = 2$, $L_p = L = 5$, $\sigma_{\text{g}} = 0$], [ARMA: $P = 4$, $Q = 0$, $\gamma = 0$], [RGP-VAR \cite{mei2017signal}: $P = 2$], [LMS on Graphs  \cite{di2016adaptive}: B.width $= 16$, $\mu_{\text{LMS}} = 1.3$], [RLS on Graphs \cite{di2017adaptive}: B.width $=16$, $\beta_{\text{RLS}} = 0.05$].}
\label{fig.epid_pred_pow}
    \end{subfigure}
\hfil
    \begin{subfigure}{.38\columnwidth}
\centering
\includegraphics[width=\columnwidth,trim={.45cm 0 .7cm .35cm},clip]{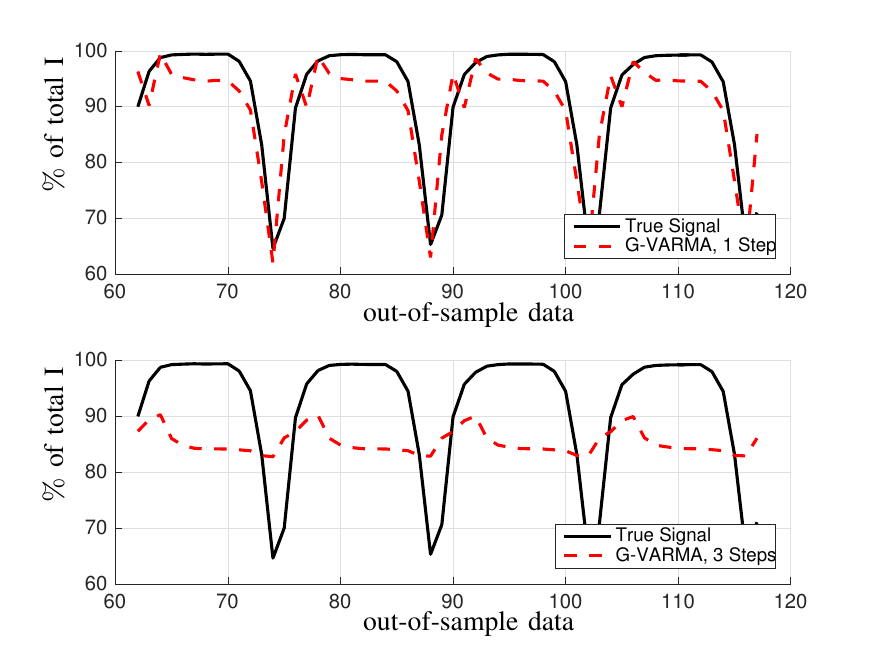}\vskip-2mm
\caption{Evolution of the infection portion on the whole graph and the G-VARMA one- and three-step ahead predictions.}
\label{fig.epid_true_sig}
    \end{subfigure}
    \end{minipage}}
    \caption{Comparison of the predictive power of the different algorithms in the three scenarios. (a)-(b) Walking dog mesh. (c)-(d) SI epidemic diffusion. The results in plots (b) and (d) are w.r.t. the out-of-sample data.}
    \label{fig.motherDogSI}
    \end{figure*}

\begin{figure*}[t]
\centering
\begin{subfigure}[h]{0.24\linewidth}
\hskip2mm\includegraphics[width=\linewidth,trim={1cm 0.5cm 1.0cm 0cm},clip]{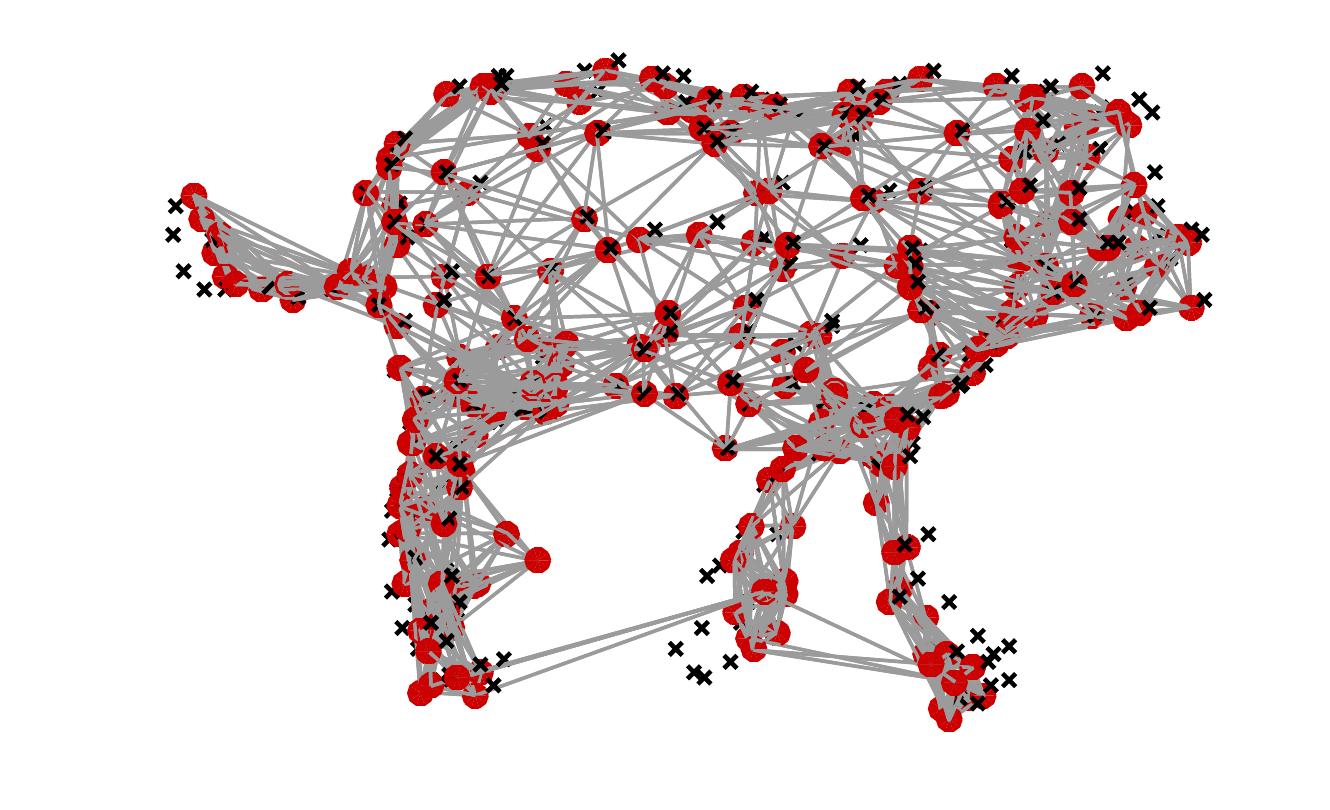}
\end{subfigure}
\begin{subfigure}[h]{0.24\linewidth}
\hskip2mm\includegraphics[width=\linewidth,trim={1cm 0.5cm 1.0cm 0cm},clip]{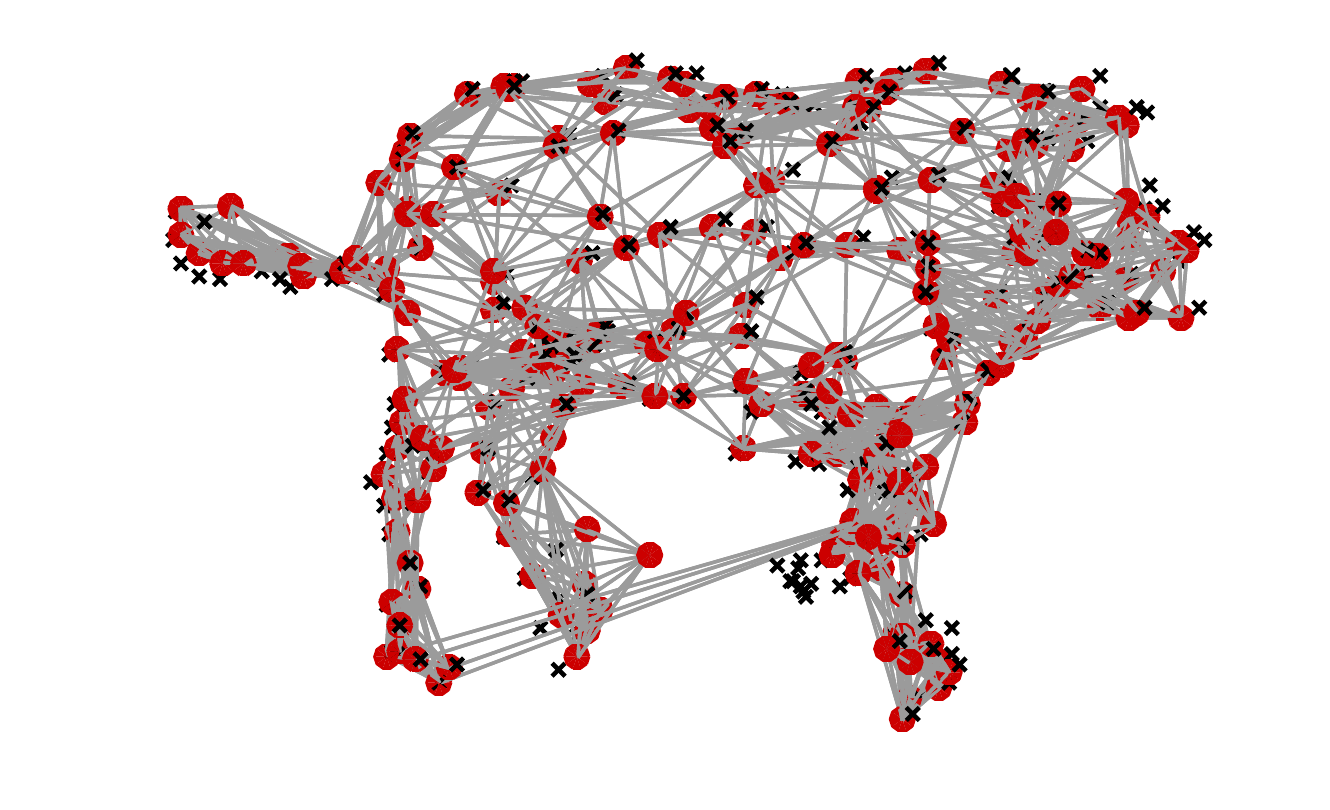}
\end{subfigure}
\begin{subfigure}[h]{0.24\linewidth}
\hskip2mm\includegraphics[width=\linewidth,trim={1cm 0.5cm 1.0cm 0cm},clip]{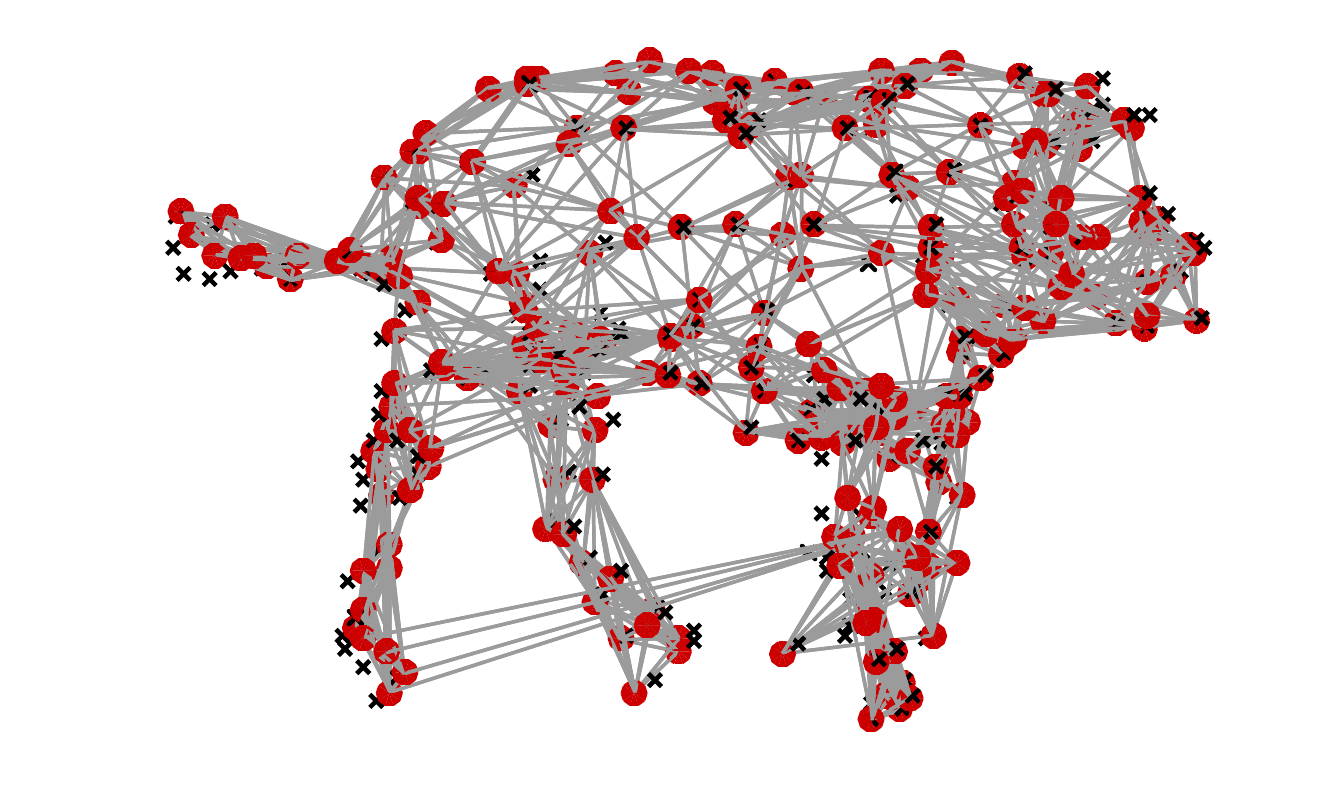}
\end{subfigure}
\begin{subfigure}[h]{0.24\linewidth}
\hskip2mm\includegraphics[width=\linewidth,trim={1cm 0.5cm 1.0cm 0cm},clip]{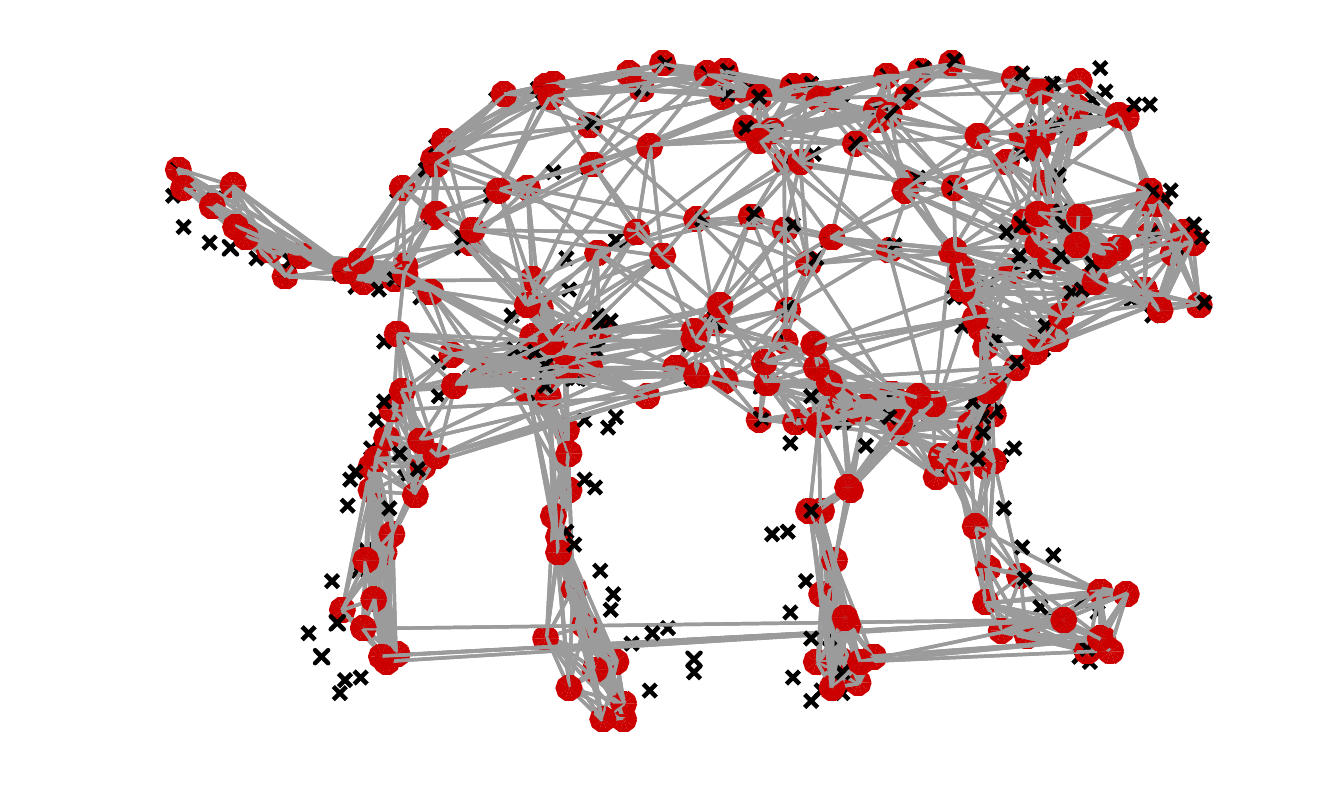}
\end{subfigure}
\caption{\rev{Four test time steps of the three step ahead G-VARMA prediction in the walking dog experiment. The red points correspond to the prediction and the black to the ground-truth. The four time steps are causal with a delay of 2 steps.}}
\label{fig.dogPos}
\end{figure*}

\subsection{Prediction performance} \label{subsec:predPerf}
We first compare the predictive power of the different algorithms. \rev{For all scenarios, the training, cross-validation, and test data consist respectively of $35\%$, $15\%$, and $50\%$.} The parameter fitting criterion is the average of the rNMSE \eqref{eq.av_rNMSE} over the five-step ahead prediction. \rev{The results for the weather data sets (Molene and NOAA) are shown in Fig.~\ref{fig.motherPred}, while those of the waling dog and epidemic diffusion in Fig.~\ref{fig.motherDogSI}.}

\vskip1mm
\noindent\textbf{Molene data set.} Fig.~\ref{fig.mol_pred_pow} shows the prediction rNMSE up to five future steps (hours) for all methods. We see that the proposed G-VARMA and GP-VAR models achieve the best performance over other approaches, although the error is small only for predictions of up to three hours. The GP-VAR falls sightly behind the G-VARMA in terms of accuracy due its more constrained form (localized graph filters).

Fig.~\ref{fig.mol_true_sig} depicts the true out-of-sample signal and the one and three-steps ahead prediction of the G-VARMA model at the red sensor in Fig.~\ref{fig.2Scen_graphs} (a). We see that the one-step ahead prediction matches well the true signal with a deviation lower than 0.5 degrees, while the three-steps ahead prediction is characterized by a larger deviation (up to $3$ degrees Celsius).

\vskip1mm
\noindent\rev{\textbf{NOAA data set.} Fig.~\ref{fig.noaa_pred_pow} shows the prediction power of the different algorithms in this scenario. We first remark that the training of standard VAR crashed continuously due to the high number of parameters involved. The graph-based techniques on the other hand, lead to trainable models due to the reduced number of parameters. These results show that the proposed methods exploit better the graph parameterization and yield the minimum performance. 
}

\rev{The disjoint ARMA compares well with the proposed algorithms and yields values closer to minimum achieved by the G-VARMA. This is not entirely surprising for two main reasons. First, the time series present a strong temporal stationarity which is known to be well exploited by ARMA models. Second, we fixed the graph as in \cite{di2017adaptive} and \cite{romero2016kernel} for a direct comparison with these methods. This choice penalizes the stationarity assumptions. Therefore, we believe that the performance of the G-VARMA and the GP-VAR can be further improved by optimizing the graph topology.
}

\rev{As a final remark, note that we used the parameters of the LMS ($\mu_{\text{LMS}} = 1.5$) and RLS ($\beta_{\text{RLS}} = 0.5$) on graphs algorithms as fixed\footnote{\rev{This is done to ensure a fair comparison with these methods.}} by \cite{di2017adaptive}. By cross-validation, we found that the optimal values are $\mu_{\text{LMS}} = 1.75$ and $\beta_{\text{RLS}} = 0.05$. These cross-validated values lead to a slightly lower rNMSE, but still significantly worse when compared with the proposed methods and to the disjoint ARMA.}

\rev{Fig.~\ref{fig.noa_true_sig} shows the one- and the three-steps ahead predictions at node $25$, which is the same used to display the tracking results of LMS, RLS, and KKF in \cite{di2017adaptive}. We observe that the one-step ahead prediction matches relatively well the true signal value, while a bigger error is observed for the three-step ahead prediction especially closer to the transition points.}

\vskip1mm
\noindent\textbf{Walking dog mesh.} The rNMSE vs number of step ahead results are shown in Fig.~\ref{fig.dog_pred_pow}. The classic multivariate VAR is not shown as it yielded unstable predictions (as the number of time series is large). Incorporating the graph in the predictive models seems to avoid these issues. In fact, all graph-based approaches give more stable predictors. Additionally, the RGP-VAR \cite{mei2017signal} yields the worst prediction due to fewer DoF. A comparison with its generalized version, i.e., the GP-VAR, shows an rNMSE close to the best performance. We note that the rNMSE gap between the proposed methods and other graph-based alternatives is larger w.r.t. the Molene data set. The spectral smoothing plays a role here for the GP-VAR since the signal evolution (i.e., the point position) is more regular along the graph dimension than temperature. This is reflected in a higher value of $\sigma_g$.

Fig.~\ref{fig.dog_true_sig} illustrates the true signal evolution along with its G-VARMA  one- and three-step ahead predictions. The selected node is the one that presents the largest peak-to-peak variability in the true signal. Compared to the Molene data set, we note a worse three-step ahead prediction, while the one-step ahead prediction is still comparable. \rev{Fig.~\ref{fig.dogPos} further shows the true dog position in four causal time steps and the three step ahead G-VARMA prediction. We can observe that the proposed model matches well the dog position in most of the nodes. Larger errors are observed around the body parts that move the most, e.g., legs and tail.}

\vskip1mm
\noindent\textbf{SI epidemic diffusion.} The results for this scenario are shown in Figs.~\ref{fig.epid_pred_pow}-\ref{fig.epid_true_sig}. Similarly to the walking dog mesh, the VAR model yields unstable results. From the results in Fig.~\ref{fig.epid_pred_pow}, all graph-based methods offer a similar rNMSE for the one-step ahead prediction\footnote{An exception is the RGP-VAR recursion \cite{mei2017signal}, which offers a worse performance. However, the focus of this approach is to learn a topology that facilitates prediction and it might suffer when the fixed graph structure differs from the learned one.}. The latter highlights the importance of the underlying topology in building prediction models, since the univariate ARMA yields a larger rNMSE than the proposed models. For more than one-step ahead predictions, we see that the proposed approaches exploit better the graph and keep yielding a smaller rNMSE than the adaptive algorithms. \rev{Nevertheless, starting from three-steps ahead, the univariate ARMA offers a similar rNMSE as the proposed methods.}

We make the subsequent remarks. First, the rNMSE is higher in this data set w.r.t. the previous two cases. We attribute this degradation to the abrupt transitions in the patient recovery instances (i.e., every $12$ days). Second, also in this data set we see that the G-VARMA does not have coefficients in the MA part. {We have found that the value of $Q$ varies depending on the amount of training data and cross-validation criterion.} For instance, the G-VARMA model with $P = 1$, $Q = 1$, $\sigma_g = 0.5$, and $\gamma = 0.44$ yields a better performance in the test data than the one shown in Fig.~\ref{fig.epid_pred_pow}, though in the cross-validation data its performance is suboptimal.
%
%
Finally, we have observed that the chosen models are characterized by a bounded memory in time. That is, only a few past realizations are useful for the prediction task. The GP-VAR model shows that there is a larger graph influence (i.e., information from further nodes) in the model. Indeed, information from nodes up to four hops away is exploited in Fig.~\ref{fig.epid_pred_pow} and in Fig.~\ref{fig.mol_pred_pow}.

To conclude, Fig.~\ref{fig.epid_true_sig} illustrates the evolution of the I-state on the whole graph and the respective G-VARMA predictions. The proposed model follows relatively well the I-state evolution for one-step ahead prediction with an error up to $4\%$, yet its performance degrades in the three-step ahead prediction\footnote{This may be improved by using other model parameters.}. {The decreased amplitude (w.r.t the ground truth) is mainly attributed to a couple outlier nodes where the I-state evolution is not predicted correctly, and, therefore, influence the overall result.}

\begin{figure}[t]
\centering
\includegraphics[width=\columnwidth,trim={.3cm 0 .7cm 0},clip]{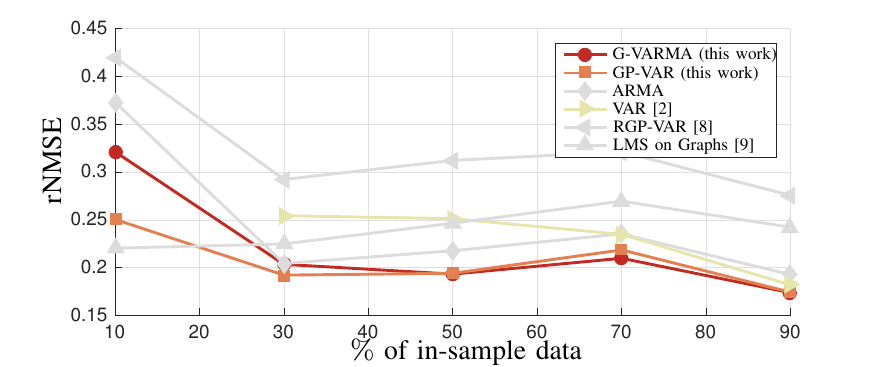}
\caption{One-step ahead prediction rNMSE versus the percentage of in-sample-data for the different algorithms in the Molene data set. The RLS on graphs \cite{di2017adaptive} is not shown here and offers a similar performance as the LMS on graphs algorithm. Note that the standard VAR could not be trained for $10\%$ in-sample data.}\vskip-.2cm
\label{fig.mol_insample}
\end{figure}

\subsection{In-sample data effect}

We now analyze the effect of the in-sample data size on prediction accuracy. For consistency with the earlier results, {we kept the cross-validation data to be $30\%$ of the in-sample data,} whereas the model parameters are selected to minimize the one-step ahead rNMSE.

Fig.~\ref{fig.mol_insample} shows the one-step ahead prediction rNMSE as a function of the in-sample data percentage for the Molene data set. We first highlight that the multivariate VAR \cite{lutkepohl2005new} is the one that suffers the most in low in-sample data regimes. For this model, we see a clear trend of rNMSE reduction when more data are used for estimating the coefficients. All graph-based methods are more robust to low in-sample data sizes since they have fewer parameters. The latter along with the spectral smoothing seems to play a role when the amount of training data is limited. The slight variation in the rNMSE for more training data is within the experimental variance. When the percentage of in-sample data is very small ($10\%$), the GP-VAR improves upon G-VARMA due to its fewer DoF.

We conclude that the proposed methods make a better use of the underlying topology and provide a good trade-off between the model complexity and the amount of training data.
As a final remark, we do not show these results for the other scenarios since the VAR \cite{lutkepohl2005new} recursion (which is the one to compare with) does not lead to stable predictors.

\begin{figure}[t]
\centering
\includegraphics[width=.9\columnwidth,trim={.3cm 0 .7cm 0},clip]{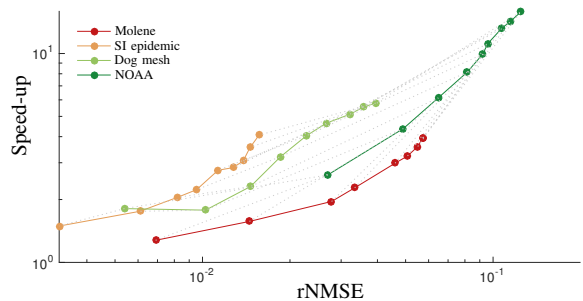}
\caption{Gained speed-up versus the loss in prediction rNMSE for the G-VARMA model estimated using the low-rank representation of the process. \rev{Both axes are in log scale to improve the visibility. For the NOAA data set, the maximum rNMSE loos is $0.13$ with a speed-up of $16$.}}
\label{fig.mol_lowRank}
\end{figure}

\subsection{Low-rank model effect}
\label{subsec_lowRank}

We now test the effects of the low-rank G-VARMA coefficient estimation described in Theorem~\ref{theo.low_rank}. Under the settings of Fig.~\ref{fig.motherPred}, we estimate the models by ignoring different portions of the process JPSD. We considered nine ignoring percentages log-spaced in the interval $[0.74\%, 5\%]$ and evaluate the rNMSE loss and the code speed-up. The code running time is averaged over $100$ iterations and the speed-up achieved using the low-rank procedure is compared w.r.t. the case where the entire JPSD is used.

Fig.~\ref{fig.mol_lowRank} shows the gained speed-up as a function of the rNMSE degradation in the one-step ahead prediction. Each marker corresponds to an increasing data ignorance percentage. The obtained results show a trade-off between prediction accuracy and computational time. We see that the most sensitive scenario to data ignorance is the NOAA data set as its NMSE loss increases the fastest, while the least sensitive scenario is the SI epidemic one. This much higher sensitivity of the NOAA data set is due to the misalignment between the used graph and the stationarity assumptions. Better results are expected in this data set with a topology that imposes a stronger joint stationarity. These observations highlight, nevertheless, the sparsity in the process JPSD which can be exploited to ease the model estimation costs.

\begin{figure}[t]
\centering
\includegraphics[width=\columnwidth,trim={.3cm 0 .4cm 0},clip]{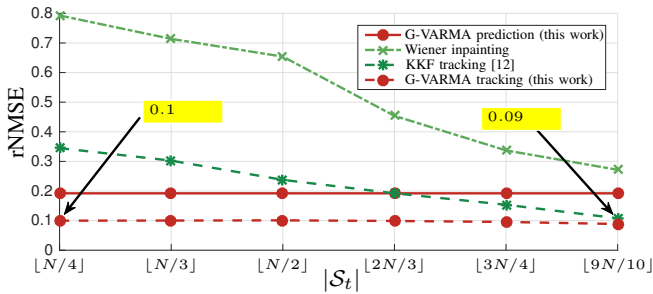}
\caption{rNMSE versus the cardinality of sampling set for different tracking algorithms in the Molene data set. The used parameters are: [G-VARMA \eqref{eq:joint_causal_model} : $P = 3$, $Q = 0$, $\sigma_{\text{g}} = 0$, $\gamma = 0$], [Wiener inpainting, $\w_t \sim \mathcal{N}(\mathbf{0}_N, 0.5\I_N)$], [KKF \cite{romero2016kernel}: $\mu_{\text{KKF}} = 10^{-7}$, $\sigma_{\text{KKF}} = 1.25$, $b_{\text{KKF}} = 0.09$].}
\label{fig.mol_track}
\end{figure}

\subsection{Tracking performance}

We here evaluate the tracking performance of the G-VARMA model on the Molene and NOAA data sets. The G-VARMA parameters are the same as the one used in Section~\ref{subsec:predPerf} with coefficients estimated following the low-rank model and ignoring $1.5 \%$ of the process JPSD. The noise power for the Wiener inpainting is cross-validated following the procedure in Fig.~\ref{fig.fit_proc} using all the $50\%$ in-sample data. \rev{For the KKF, we cross-validated the different parameters, i.e., the KKF kernel weight ($\mu_{\text{KKF}}$), the diffusion coefficient ($\sigma_{\text{KKF}}$), and the edge propagation weight ($b_{\text{KKF}}$) using all the $50\%$ in-sample data in the Molene data set. Instead, for the NOAA data set we used the parameters proposed in \cite{romero2016kernel}.} During the training process, $75\%$ of the nodes are sampled uniformly at random and the considered score is the tracking rNMSE. To the G-VARMA tracker, a small artificial noise is added to the out-of-sample data to avoid numerical issues when inverting $\H_t$. In the test phase, we average the rNMSE over $100$ random node selections and consider $\bar{\Scal}_t = \Vx$.

\vskip1mm
\textbf{Molene data set}. Fig.~\ref{fig.mol_track} depicts the average tracking rNMSE as a function of the size $|\Scal_t|$ of the observation set. The proposed method is seen to improve substantially the accuracy of the G-VARMA predictor and outperforms the other approaches even with $N/4 = 8$ collected samples at time $t$. Wiener inpainting provides suboptimal predictions as it does not exploit the historical data. These results show also the benefits of the G-VARMA predictors over the Wiener inpainting and the KKF approach.

\begin{figure}[!t]
\centering
    \begin{subfigure}{\columnwidth}
\centering
\includegraphics[width=\columnwidth,trim={.3cm 0 .4cm 0},clip]{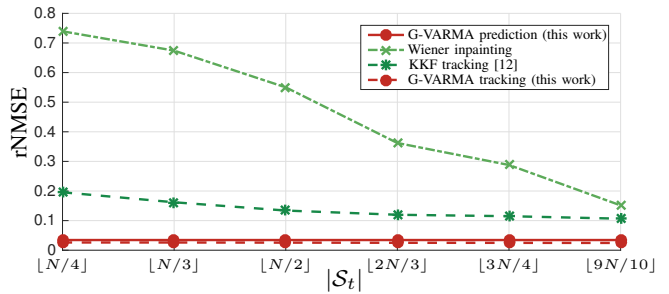}
\caption{Comparison of the different methods.}
\label{fig.noaa_track}
    \end{subfigure}
    \caption{rNMSE versus the cardinality of sampling set for different tracking algorithms in the Molene data set. The used parameters are: [G-VARMA \eqref{eq:joint_causal_model} : $P = 3$, $Q = 0$, $\sigma_{\text{g}} = 0$, $\gamma = 0$], [Wiener inpainting, $\w_t \sim \mathcal{N}(\mathbf{0}_N, 0.5\I_N)$], [KKF \cite{romero2016kernel}: $\mu_{\text{KKF}} = 10^{-7}$, $\sigma_{\text{KKF}} = 1.8$, $b_{\text{KKF}} = 0.01$]. A zoomed in version of the G-VARMA tracking algorithm is provided in the Appendix.}
    \label{fig.motherTracNoaa}
    \end{figure}

\vskip1mm
\rev{
\textbf{NOAA data set}. The tracking results for the NOAA data set are shown in Fig. \ref{fig.noaa_track}. The latter enforces the behavior of the different methods observed for the Molene data set. The Wiener inpainting still offers the worse performance as it does not explore the temporal dimension. Similarly, both the G-VARMA prediction and tracking offer a lower rNMSE than the KKF \cite{romero2016kernel}. The performance of the KKF can be slightly improved by using cross-validation (validated parameters $\mu_{\text{KKF}} = 10^{-6}$, $\sigma_{\text{KKF}} = 1.5$, $b_{\text{KKF}} = 0.15$).}

\rev{In Fig. \ref{fig.noaaTrackZoom} (see the Appendix), we zoomed in the performance of the proposed G-VARMA tracking methods. A larger cardinality of the sampling set leads to small reduction of the rNMSE. We mainly attribute this minimal improvement to the mismatch between the data behavior and the G-VARMA model. Further reduction of the rNMSE can be obtained by selecting the $|\mathcal{S}_t|$ in a sparse sensing fashion \cite{chepuri2016sparse} and by introducing seasonality into the G-VARMA model \cite{brockwell2002introduction}.}

\section{Conclusion}
\label{sec.concl}

%
The purpose of this work was to determine multivariate models for forecasting the temporal evolution of time series on graphs. 
%
By leveraging the (approximate) joint time-vertex stationary of the time series, we reduced the DoFs for the VAR and VARMA models by designing their coefficients in the graph spectral domain. The latter decomposition allowed us to fit the model parameters with well-known univariate techniques by tackling both the computational and stability issues present in classical VAR and VARMA models. To further reduce the estimation costs, we proposed an optimal low-rank representation for the time series. Finally, a sub-graph tracker was introduced when measurements form a subset of nodes become available.

The findings clearly indicated that the graph structure plays a major role in easing the computational complexity of high dimensional multivariate models. This role is further enhanced when the amount of training data is limited. 
Several numerical experiments showed the benefits of the proposed methods over the classical multivariate forecasting techniques and other the state-of-the-art graph-based approaches.

Overall, this work strengthens the idea that practical benefits could be achieved in the time series forecasting if the underlying data connections are exploited. Further work is needed in two main directions. First, it is important to understand the implications of the stationary assumption for the task of prediction. Second, it is useful to analyze the effects of the graph structure in the prediction accuracy.


\section*{Appendix}

%

\section*{Proof of Proposition~\ref{prop:mse_1step}}
\label{sec_appPredMSE}

By substituting the expression of $\tilde{\x}_{t}$ \eqref{eq:predictor1} and $\x_t$ \eqref{eq:joint_causal_model} into the the one-step ahead predictor $\text{MSE} = \Exp\left[\|\tilde{\x}_{t} -\x_t\|_2^2\right]$, we have
%
\begin{align}
\label{eq.MSE_dummy2}
\begin{split}
\text{MSE} =  \Exp\left[\left\|\sum_{q = 1}^Q b_q(\L_\Gr) \left( \x_{t-q} - \tilde{\x}_{t-q}  \right) - \sum_{q = 0}^Q b_{q}(\L_\Gr) \, \eps_{t-q}\right\|_2^2	\right].
\end{split}
\end{align}
Then, by substituting $\eps_{t-q} = \x_{t-q}-\tilde{\x}_{t-q}$ in \eqref{eq.MSE_dummy2} we obtain
\begin{align}
\label{eq.MSE_dummy3}
\begin{split}
\text{MSE} =  \Exp\left[\|b_0(\L_\Gr)\eps_t\|_2^2		\right],
\end{split}
\end{align}
which concludes the derivation of \eqref{eq:MSE1Step}. The MSE \eqref{eq.MSE_dummy3} depends only on the unknown innovation at time $t$, which corresponds to the smallest achievable MSE for the given set of coefficients $\{a_p(\L_\Gr)$, $b_q(\L_\Gr)\}$. This concludes the proof. $\qed$

\section*{Proof of Proposition~\ref{prop.decoup}}
\label{sec_appDecoupMSE}

By regrouping the terms containing $\x_\tau$ on the left hand-side of \eqref{eq:joint_causal_model}, the $n$th element rotated by the unitary matrix $\U_\Gr^\herm$ is
\begin{align}\label{eq.proof1}
\begin{split}
\left[\U_\Gr^\herm\sum_{p = 0}^P a_p(\L_\Gr) \x_{t-p}\right]_{n} &= \left[\sum_{p = 0}^P \left[a_p(\LAM_\Gr)\right]_{nn} \U_\Gr^\herm \x_{t-p}\right]_{n}\\
&= \sum_{p = 0}^P  \left[a_p(\LAM_\Gr)\right]_{nn} \hat{x}_{t-p,n},
\end{split}
\end{align}
where $a_0(\L_\Gr) = a_p(\LAM_\Gr) = \I_N$
Similarly, the $n$-th element of the remaining right hand-side of \eqref{eq:joint_causal_model} rotated by $\U_\Gr^\herm$ is
\begin{align}\label{eq.proof2}
\begin{split}
\left[\U_\Gr^\herm\sum_{q = 0}^Q b_q(\L_\Gr) \eps_{t-q}\right]_{n} &= \left[\sum_{q = 0}^Q b_q(n)\U_\Gr^\herm \eps_{t-q}\right]_{n}\\
&= \sum_{q = 0}^Q \left[b_q(\LAM_\Gr)\right]_{nn} \, \hat{\varepsilon}_{t-q,n}.
\end{split}
\end{align}
Then, the claim \eqref{eq:prop2} is obtained by direct substitution of \eqref{eq.proof1}-\eqref{eq.proof2} into \eqref{eq:joint_causal_model}. $\hfill\blacksquare$

\section*{Proof of Theorem~\ref{theo.low_rank}}
\label{sec_appLowRnkTheo}

Let us define $\A = \U(\I_N-\D_\Scal)\U^\herm$. Following the Eckart–Young–Mirsky Theorem~\cite{eckart1936approximation,markovsky2008structured}, the expected approximation error is
\begin{align}
\label{eq.proof1}
\begin{split}
\Exp\!\left[\!	\left\|	\X_{1:T}\!-\!\tilde{\X}_{\U,\Scal}	\right\|^2_F	\!\right] \!\!&=\! \Exp\!\left[\!\left\|\A\X_{\!1:T}	\right\|^2_F\!\right] \!=\! \trace\!\left(\!\A\Exp[\X_{\!1:T}\X_{\!1:T}^\herm]\A^\herm	\right)\!\!.
\end{split}
\end{align}
Then, from Theorem 2 of~\cite{perraudin2016towards}, where for each time instant $t$, the graph signal $\x_t$ is JWSS stationary with covariance $\bSigma_t = \Exp[\x_t\x_t^\herm] = \U_\Gr g_t^2(\LAM_\Gr)\U_\Gr^\herm$ implying that
\begin{align}
\begin{split}
\bSigma_\Gr &= \Exp[{\X_{1:T} \X_{1:T}^{\herm}}] =  {\sum_{t=1}^T\Exp[\x_t\x_t^\herm}] = \sum_{t=1}^T \bSigma_{x}\\
&= \U_\Gr g^2(\LAM_\Gr)\U_\Gr^{\herm}
\end{split}
\end{align}	
%
where $g^2(\LAM_\Gr) = \sum_{t = 1}^Tg_t^2(\LAM_\Gr)$ are reordered such that $g(\lambda_1) \ge g(\lambda_2) \ge \ldots \ge g(\lambda_N)$. By substituting of $\bSigma_\Gr$ into \eqref{eq.proof1} we get
\begin{align}
\label{eq.proof2}
\begin{split}
\Exp\!\!\left[\!	\left\|	\X_{\!1:T}\!-\!\tilde{\X}_{\U,\Scal}	\right\|^2_F	\!\right] \!&=\! \trace\! \left(\!\A\bSigma_\Gr\A\!{^\herm}	\right) \!=\! \left\|\bSigma_\Gr^{1/2} \!\!-\! \U\D_\Scal\U^\herm\bSigma_\Gr^{1/2}		\right\|_F^2\\
&= \left\|g(\LAM_\Gr) -  	\U\D_\Scal\U^\herm g(\LAM_\Gr)	\right\|_F^2.
\end{split}
\end{align}
Setting then $\B = \U\D_\Scal\U^\herm$, \eqref{eq.proof2} becomes
\begin{align}
\label{eq.proof3}
\begin{split}
&\Exp\left[	\left\|	\X_{1:T}-\tilde{\X}_{\U,\Scal}	\right\|^2_F	\right] = \sum_{i = 1}^N\left|g(\lambda_i) - [\B]_{ii}g(\lambda_i)	\right|^2 \\
&\qquad\qquad+ \sum_{i \neq j}\left|[\B]_{ij}g(\lambda_i)	\right|^2 \ge  \sum_{i = 1}^N\left|g(\lambda_i)	\right|^2\left|1-[\B]_{ii}	\right|^2\\
& \overset{(a)}{\ge} \sum_{i = K+1}^N|g(\lambda_i)|^2  = \Exp\left[	\left\|	\X_{1:T}-\tilde{\X}_{\U_{\Gr},\Scal^\star}	\right\|^2_F	\right],
\end{split}
\end{align}
where in the third step we use the fact that $[\B]_{ii} \le 1$ and is exactly 1 at most $K$ times. The last expression shows that the global minimum is achieved for $\tilde{\X}_{\U_{\Gr},\Scal^\star}$ with $\Scal$ containing the $K$ largest components of $g(\lambda)$. In fact, $(a)$ is independent on the choice of $\B$, while the last equality holds only for $\U = \U_{\Gr}$. Any other rotation will lead to a different $\B$ thus a larger error. $\hfill\blacksquare$

\section*{Zoomed in figures}
\label{sec_appZoom}

\begin{figure}[t]
\centering
\includegraphics[width=\columnwidth,trim={.3cm 0 .4cm 0},clip]{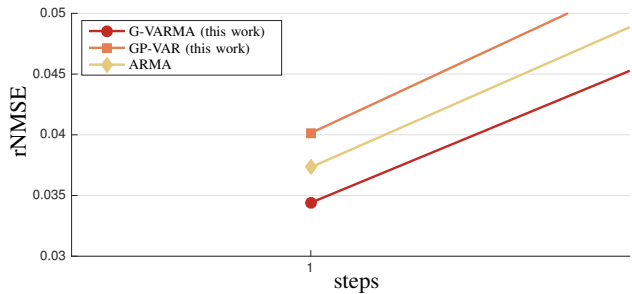}
\caption{Zoomed in version of Fig.~\ref{fig.noaa_pred_pow} that shows the minimal differences between the proposed G-VARMA, GP-VAR, and the disjoint ARMA approach. We can see that the G-VARMA model slightly improves the disjoint ARMA approach and this gap increases for more steps ahead predition.}
\label{fig.noaaZoom}
\end{figure}

\begin{figure}[!t]
\centering
\includegraphics[width=\columnwidth,trim={.1cm 0 .4cm 0},clip]{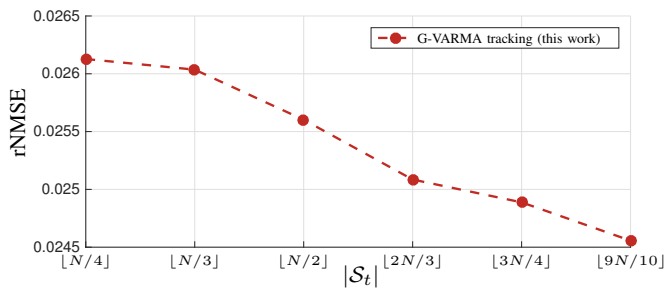}
\caption{Zoomed in version of Fig.~\ref{fig.motherTracNoaa} that shows the reduction of the rNMSE for the G-VARMA tracking algorithm when more nodes are sampled at time $t$.}
\label{fig.noaaTrackZoom}
\end{figure}

This section presents the zoomed in versions of Fig.~\ref{fig.noaa_pred_pow} in Fig.~\ref{fig.noaaZoom} and Fig.~\ref{fig.motherTracNoaa} in Fig.~\ref{fig.noaaTrackZoom}.

\bibliographystyle{IEEEtran}
\bibliography{MyPaperCollection,references}

\end{document}